\documentclass[11pt]{article}
\usepackage{tikz}
\usepackage{subcaption}
\usepackage{eurosym}
\usepackage{amsthm}

\newtheorem{remark}{Remark}
\usepackage{amsmath}
\usepackage{amssymb}
\usepackage{mathtools}
\usepackage{graphicx}
\usepackage{setspace}
\usepackage{sectsty}
\usepackage[authoryear]{natbib}
\usepackage[margin=1in]{geometry}
\usepackage{color}
\IfFileExists{bbm.sty}{\usepackage{bbm}}{}
\usepackage{booktabs}
\usepackage{pdflscape}
\usepackage{dcolumn}
\usepackage{threeparttable}
\IfFileExists{siunitx.sty}{\usepackage{siunitx}}{}
\usepackage{array}
\usepackage{appendix}
\usepackage{enumitem}
\usepackage{placeins}
\usepackage{fancyhdr}
\usepackage[breaklinks=true]{hyperref}
\usepackage{cases}
\usepackage{etoolbox}
\usepackage{titlesec}
\usepackage{ragged2e}
\usepackage{lmodern,textcomp}
\usepackage[utf8]{inputenc}
\usepackage{longtable} 
\usepackage{import}

\theoremstyle{plain}                     

\theoremstyle{definition}                

\begin{document}

    \hypersetup{pageanchor=false}
\begin{titlepage}
\renewcommand{\baselinestretch}{1}
	\title{
		\sc{
			\Large{
            Uncertain and Asymmetric Forecasts
			}%
		}%
	\thanks{I am grateful to Olivier de Bandt, Jean Barth\'{e}lemy, Agn\`{e}s Benassy-Qu\'{e}r\'{e}, Antoine Camous, Laurent Clerc, Jose Fellmann, David Gauthier, Christoph Grosse-Steffen, Sophie Guilloux-Nefussi, St\'{e}phane Lhuissier, Magali Marx, Adrian Penalver, Arthur Saint-Guilhem, Manuel Schick and Elias Wolf for helpful comments and discussions. The views expressed in this paper are solely those of the author and do not necessarily reflect the views of his past, present, or future employers. Replication code for every figure and table, and for the measures themselves, is available at \url{https://github.com/skimeur/normalized-uncertainty}.}
	}
\author{
\textbf{Eric Vansteenberghe}%
\thanks{
Banque de France and Universit\'{e} Paris~1 Panth\'{e}on-Sorbonne. E-mail: \href{mailto:eric.vansteenberghe@banque-france.fr}{\nolinkurl{eric.vansteenberghe@banque-france.fr}}
}
}

	\date{}

	\maketitle

\abstract{\noindent
Survey density forecasts are summarized by their second and third moments, read as uncertainty and as the balance of risks. Neither can be read on its own. In the ECB Survey of Professional Forecasters the variance of an individual inflation density rises with the distance of that forecaster's central forecast from the official target, flat below it and rising above, so that raw dispersion mixes belief imprecision with the arithmetic of the level; and a third moment computed over a handful of bins is too noisy to identify directional risk unless it is read together with the location of the density. This paper builds, from the reported bins up, two measures that repair these defects. Normalized Uncertainty divides a density's standard deviation by the one the fitted variance--distance envelope predicts at the observed distance. Asymmetry Coherence retains directional risk only where the sign of the reported asymmetry agrees with the sign of the median's deviation from the target, weighted by that agreement. Both are carried to growth densities, where the reference is estimated rather than announced; to a simulation in which the latent objects are known; and to the US Survey of Professional Forecasters.
\label{abstract}
}



\thispagestyle{empty}
\end{titlepage}
\hypersetup{pageanchor=true}

    \section{Introduction}
\label{sec:intro}

Raw higher moments of forecast distributions do not necessarily identify the economic objects they are commonly taken to measure. When inflation expectations move away from a policy anchor, raw forecast variance increases mechanically---even if genuine belief imprecision does not---so that what is often interpreted as ``uncertainty'' partly reflects the level of inflation itself \citep{friedman1977nobel, ball1992does}. Raw skewness poses a related but different challenge: it is difficult to estimate precisely from forecast data or option prices with limited strikes. The first moment provides useful discipline for interpreting the third moment and identifying directional risk. Accordingly, the second raw moment embeds a first-moment component large enough to distort inference, whereas the third moment becomes more informative once disciplined by the first moment.

This paper proposes two corrections that recover the objects of interest by appropriately reinterpreting their relationship with the first moment. \emph{Normalized Uncertainty} (NU) disentangles the first and second moments by removing the predictable level component from raw dispersion through a variance-stabilizing transformation anchored in the distance-to-target relationship, thereby recovering genuine belief imprecision. \emph{Asymmetry Coherence} (AC), by contrast, re-links the first and third moments by retaining only the component of asymmetry that is coherent with the central forecast, yielding an operational measure of directional risk and the balance of risks. Both corrections are grounded in micro-founded mechanisms and validated using individual density forecasts from the ECB Survey of Professional Forecasters (ECB-SPF). The mechanism behind the variance--distance relationship, and the
one-sided form it takes around an announced target, are developed in
\citet{vansteenberghe2026tolerable}.

\paragraph{What this paper provides.}
Five things. First, the construction of Normalized Uncertainty from the survey histograms:
the moments read off the reported bins, the trimming the open tails require, the two-arm
variance--distance envelope with its estimates on the 1999--2026 panel, and
two operational calibrations of the normalization --- one with the envelope estimated, one
with it set to unity --- reported side by side. Second, the
construction of Asymmetry Coherence, a bounded statistic that reads the third moment
through the first and so extracts a directional signal from densities whose individual
asymmetry is too noisy to be read on its own. Third, the growth counterparts, Normalized
Growth Uncertainty and Asymmetry Coherence of Growth, built against potential growth rather
than an announced target, together with an orthogonalization of inflation and growth
uncertainty carried out forecaster by forecaster. Fourth, the rules for carrying the
constructions to another survey --- compare shapes and within-source ratios, never slopes
across sources, and control for a change of histogram grid --- illustrated on the US Survey
of Professional Forecasters, where a halving of bin widths in 2014, and not the 2012
announcement of a numerical objective, accounts for a break a naive reading attributes to
policy. Fifth, a simulation in which the latent objects are known, which checks that the
corrected measures recover them where the raw moments do not.

The remainder of the paper is organized as follows.
Section~\ref{sec:data} fixes the survey, the grid and the moment conventions every measure
below is computed under.
Section~\ref{sec:nu} builds Normalized Uncertainty and Section~\ref{sec:ac} Asymmetry
Coherence.
Section~\ref{sec:growth} carries both to growth densities and orthogonalizes the two
uncertainty measures.
Section~\ref{sec:other_surveys} carries them to the US survey.
Section~\ref{sec:simulation} validates them by simulation, and
Section~\ref{sec:conclusion} concludes.

\subsection{Related literature}
\label{subsec:related_literature}

This paper connects to several strands of the literature.

\paragraph{Moments of survey densities.}
The practice of reading uncertainty off reported histograms goes back to
\citet{zarnowitz1987consensus}, who separate the average individual uncertainty from
disagreement about the central forecast, and to \citet{engelberg2009comparing}, who fit
distributions to the reported bins and document the discipline the exercise requires.
\citet{rich2010relationships} and \citet{abel2016measurement} ask whether the resulting
uncertainty co-moves with the level of expectations and report that it does not; the
reading offered here is that on a sample in which expectations never leave the anchor, the
question is asked where the answer cannot be identified.
Special surveys run by the ECB indicate that the reported densities are predominantly
judgment-based \citep{ECB2018SpecialSurvey}, which is what makes their higher moments
informative about beliefs rather than about a common model.

\paragraph{Measuring macroeconomic uncertainty.}
Since uncertainty is not directly observable, economists have developed diverse proxies.
News-based indices count uncertainty-related keywords in newspapers \citep{baker2016measuring, ahir2022world}.
Market-based measures exploit option-implied volatility to infer the price of aggregate risk.
Statistical approaches estimate the expected volatility of the unforecastable component of many macro variables \citep{jurado2015measuring}.
Survey-based approaches, pioneered by \citet{zarnowitz1987consensus}, exploit histogram forecasts to distinguish \emph{average individual uncertainty} from \emph{disagreement} \citep{engelberg2009comparing}.
A common feature across all these approaches is that the resulting measures tend to co-move with first-moment movements---option-implied volatility depends on moneyness, GARCH variances absorb regime shifts, and survey dispersion widens when expectations drift from anchors.
Normalized Uncertainty addresses this level dependence directly: it applies a variance-stabilizing transformation that removes the predictable component of dispersion, yielding a measure of genuine belief imprecision that is portable across data sources.

\paragraph{Variance-stabilizing transformations.}
The device used here to remove a mean--variance dependence is classical: when the variance
of a random variable is a known function of its mean, dividing by the square root of that
function stabilizes it \citep{anscombe1948transformation, nelder1972generalized}, the same
family as the power transformation of \citet{box1964analysis}. What the survey data add is
that the function is not postulated but estimated, and that its argument --- the distance of
the forecast from an announced anchor --- is observed.

\paragraph{Asymmetry, directional risk, and the balance of risks.}
Skewness is economically important across multiple domains.
In asset pricing, \citet{kraus1976skewness} establish that coskewness is a priced risk factor; \citet{harvey2000conditional} find that systematic skewness commands a risk premium; and \citet{barberis2008stocks} show that idiosyncratic skewness can be overpriced under prospect theory.
In macroeconomics, \citet{neftci1984economic} demonstrates that many macro series reject symmetry, \citet{barro2006rare} shows that tail asymmetry is central to disaster-risk premia, and \citet{guvenen2014nature} documents that labor income changes exhibit countercyclical negative skewness.
For inflation specifically, \citet{ball1995relative} establish in a menu-cost framework that aggregate inflation is positively related to the skewness of relative price changes; \citet{kitsul2013economics} extract option-implied inflation densities that display pronounced time-varying skewness; and \citet{bianchi2022belief} document that shifts in the skewness of SPF inflation expectations are informative for subsequent movements in inflation and output.
Despite this rich literature, a common shortcoming is that raw asymmetry---whether derived from survey histograms, option-implied densities, or model-based distributions---is often interpreted as a fully informative measure of directional risk.
Yet the third moment is notoriously difficult to measure precisely in the settings most often used in practice: survey densities are binned and coarse, option-implied densities are recovered from a limited range of strikes and maturities, and model-based distributions are subject to specification error.
Under these conditions, raw skewness can be too noisy to stand on its own as a reliable indicator of the direction of future risks.
The point is therefore to discipline its interpretation.
The AC index does so by using the first moment to filter the third moment, retaining only the component of asymmetry that is coherent with the central forecast and therefore more likely to carry economically meaningful directional information.
It thus offers an operational statistical formalization of the ``balance of risks''---a notion central to monetary policy communication and risk management, but one that has lacked a standard empirical counterpart.

\subsection{Why dispersion moves with the distance from target}
\label{sec:distance_volatility}

The empirical uncertainty literature relies on four broad classes of proxies, all of which produce measures that co-move with first-moment movements.
\emph{Realized-volatility approaches}---beginning with \citet{engle1982autoregressive}'s ARCH framework and its GARCH extensions \citep{grier1998inflation}---estimate conditional variance from squared residuals, which mechanically inherit regime shifts in the inflation level.
\emph{State-space models}, most prominently the unobserved-components stochastic-volatility (UCSV) framework of \citet{stock2007has}, decompose inflation into trend and transitory components with time-varying volatility; the estimated stochastic volatility tracks the scale of shocks but does not separate level-driven from belief-driven variation.
\emph{News-based indices} \citep{baker2016measuring, ahir2022world} count uncertainty-related keywords in newspapers; these co-move with inflation episodes because the same macroeconomic conditions that move expectations also generate news coverage.
\emph{Survey-based measures} \citep{zarnowitz1987consensus, engelberg2009comparing} extract the variance of individual forecast distributions from histogram data, providing the most granular proxy---but one that is again mechanically tied to the level of expectations.

The contamination is quantitatively large. In the ECB Survey of Professional Forecasters, regressing raw inflation uncertainty (the cross-sectional mean of individual SPD variances) on the two arms of the distance of the consensus forecast from the 2\% target yields $R^2=0.697$ (Table~\ref{tab:arms}); on the 109 rounds through 2026Q2 it is $0.705$, the $71\,\%$ the companion paper reports \citep{vansteenberghe2026tolerable}.
Seven-tenths of the variation in what is routinely called ``inflation uncertainty'' is thus explained by a single first-moment object, once that object is allowed its one-sided shape.
Any regression that uses raw dispersion as an uncertainty proxy is therefore mechanically contaminated by this level dependence---a concern first noted by \citet{gordon1971steady} and \citet{okun1971mirage}, formalized by \citet{cukierman1979differential}, but not systematically corrected in the subsequent literature.

A robust empirical regularity of survey-based forecast distributions is
that dispersion increases when the expected value of the target variable
deviates from a policy anchor
\citep{friedman1977nobel,ball1992does, brunner1993higher,giordani2003inflation,mankiw2003disagreement,lahiri2006modelling,hartmann2022inflation}.

\paragraph{The shape.}
The dependence is not symmetric. On the euro-area panel the below-target arm is flat and
statistically indistinguishable from zero, while the above-target arm is steep: measured
uncertainty is priced when the forecast sits above the announced target and is not priced
when it sits below (Section~\ref{sec:nu_envelope}). The one-sided law, with its kink at the announced target,
is derived in
\citet{vansteenberghe2026tolerable}, of which this paper is the measurement companion.

\paragraph{Why it happens.}
Here only mention \citet{vansteenberghe2026tolerable}, then:
The object being filtered is never observed. The announced target is public; the
deviation from it a central bank will accept before acting is not, and it is never
published --- it can only be read off the policy it eventually produces. A panel of experts
each reporting an assessment of a parameter none of them observes, with the cross-section of
those reports the only evidence available, is a configuration that is not confined to
monetary policy: \citet{vansteenberghe2026insurance} meets it in insurance supervision,
where the tail parameter of a loss distribution cannot be estimated directly because the
events that would reveal it are rare, and recovers it from the direction in which expert
assessments move once the underlying risk has shifted. How the learning proceeds here ---
how beliefs about a latent tolerance converge, and how fast --- is taken up neither in this
paper nor in the companion, which treats the tolerance as fixed and shows that it is not
identified from what a forecaster sees.
    \section{The survey data and the moments of a density forecast}
\label{sec:data}

Everything in this paper is computed from reported probability distributions, so the
conventions used to turn a histogram into a moment are part of the measure and not a
detail of implementation. This section fixes them once: the survey, the calendar, the
grid on which forecasters answer, the moments read off that grid, and the reason every
statistic below is formed forecaster by forecaster before it is averaged. Sections~%
\ref{sec:nu} to~\ref{sec:simulation} use these conventions without
restating them.

\subsection{The ECB Survey of Professional Forecasters}
\label{sec:spf}

\paragraph{Data: the ECB Survey of Professional Forecasters.}
The data are the ECB-SPF, a quarterly survey initiated in 1999 that collects expectations on key euro-area macroeconomic indicators---including inflation (Harmonized Index of Consumer Prices, HICP), real GDP growth, and unemployment---from approximately 90 stable forecasting institutions such as banks and research institutes.
Participants provide quantitative forecasts for multiple horizons (one, two, and five years ahead, plus the next quarter) using both rolling and calendar horizon approaches.
A distinctive feature is its large sample size (around 60 anonymous respondents quarterly) and the inclusion of subjective probability distributions (SPDs), enabling quantification of forecast uncertainty through assigned probabilities across predefined intervals.
These SPDs are predominantly judgment-based, as indicated by special ECB surveys \citep{ECB2018SpecialSurvey}, and provide rich insights into forecasters' central tendencies and associated uncertainty \citep{garcia2003introduction}.
Individual forecasters allocate positive probability mass to, on average, nearly six distinct bins, yielding a mean normalized entropy of about 0.5.
This level of bin usage indicates that respondents meaningfully exploit the probabilistic format rather than concentrating mass on a single interval.
Hence, the ECB-SPF SPDs are sufficiently informative to support the construction of dispersion-based uncertainty measures (NU) as well as asymmetry-based indicators (AC), and cannot be reduced to degenerate or purely point-forecast representations.

\paragraph{Survey calendar and date alignment.}
The ECB-SPF is conducted four times a year: the Q1 wave is fielded in January--February, the Q2 wave in April--May, the Q3 wave in July--August, and the Q4 wave in October--November; results are typically published by the ECB six to eight weeks after the survey period closes.  The microdata files distributed by the ECB label each SPD with the \emph{target period}---defined as the calendar quarter in which the one-year-ahead forecast horizon ends.  For instance, the Q1-1999 survey (fielded January--February 1999) targets December 1999 as the endpoint of its twelve-month window, but in regression analyzes that join the NU to monthly or quarterly financial series, it is used as of the field date (\texttt{1999-01-01}), to reflect the uncertainty and asymmetry in professional forecasts at that time.

\subsection{The histogram grid}
\label{sec:grid}

Forecasters do not report a density; they report probability mass over a fixed grid of
intervals, and the grid has changed twice in ways that matter for a second moment. Through the 2024Q3 round the
inflation grid is fourteen bins: twelve half-point interior bins from $-1$ to $5\,\%$, and an
open bin at each end. The survey has used several open top bins over its history ---
$[3.5,\infty)$, $[4,\infty)$ and $[5,\infty)$ --- and they are pooled here into a single top
bin. From the 2024Q4 round the questionnaire was revised to thirteen bins whose interior
grid is shifted by a quarter point, with the open top bin at $[4.8,\infty)$. Two consequences run through the paper. First, a variance is
only as well measured as the tails are closed: when a forecaster places substantial mass
in the top open bin, where that mass sits inside the interval is the analyst's assumption
and not the forecaster's report, and the measured second moment inherits the assumption.
Outside the 2022--2023 episode the top bin carries a negligible share of the mass and the
point is moot; inside it the share is large enough that the level of measured uncertainty
in those rounds should be read as an order of magnitude rather than a measurement
(Section~\ref{sec:nu_envelope}). Second, a change of grid changes a measured
variance mechanically, with no change in beliefs; Section~\ref{sec:other_surveys} shows
what that does to a cross-source comparison when it is not controlled for.

Four conventions close the construction, and they are the ones coded in the panel builder.
Each reported probability is renormalized so that a forecaster's bins sum to one within the
grid regime that round belongs to. The two open tails are closed on the realized range: the
top bin becomes $[5,\,10.62]$ before the revision and $[4.8,\,10.62]$ after it, $10.62\,\%$
being the highest year-on-year euro-area inflation realized over the sample; the bottom bin
is closed at $-1$ (at $-0.75$ after the revision), which is below the lowest realized value,
$-0.62\,\%$, so that the bottom tail carries a point mass at its edge rather than an
interval. Each bin's mass is then represented by the midpoint of its closed interval for the
mean and the variance, while the quartiles interpolate linearly inside the bin where the
cumulative probability crosses. Finally, a range restriction. The round-level estimates exclude any survey round whose
consensus forecast lies outside $[-1,5]$, because the variance of a density centred near zero is
compressed by the grid and the variance of one centred above the top bin's lower edge is set by
where that bin is closed rather than by beliefs. One round falls out, 2022Q4, whose consensus
stood at $5.01$. The $17$ individual densities with means above $5$ that sit in other rounds,
all between 2022Q1 and 2023Q3, stay in their rounds' averages; the fits that pool
forecaster-rounds apply the same bounds to each density instead.%
\footnote{Applied to individual densities rather than to rounds, the rule leaves the headline
envelope where it is: through 2026Q2 the two arms explain $70.6\,\%$ of the variation in the
round mean of the individual variances, against $70.5\,\%$ under the round rule, with an upper
arm of $0.912$ against $0.855$.}
Running this construction from the raw survey rounds reproduces the estimates of
Section~\ref{sec:nu_envelope} exactly.

\begin{figure}[tbp]
  \centering
  \includegraphics[width=\linewidth]{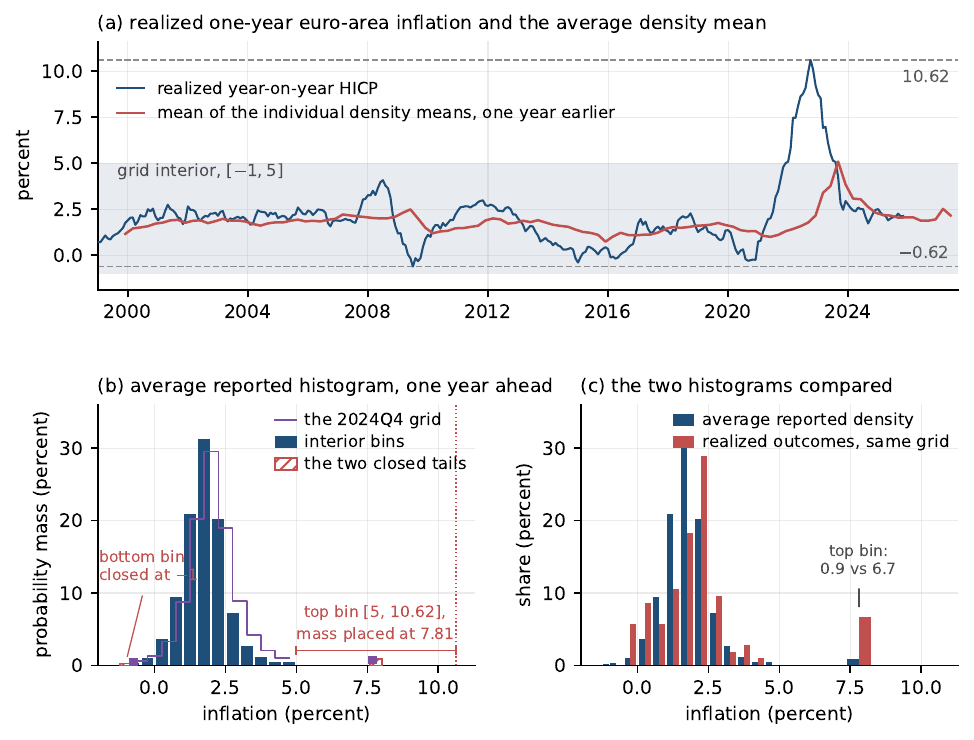}
  \caption{\textbf{The histogram grid, the realized series and the reported densities.}
  \emph{Panel (a):} realized year-on-year euro-area HICP inflation and the round mean of the
  individual one-year-ahead density means, the latter plotted at the date its forecast refers to;
  the shaded band is the interior range of the grid, $[-1,5]$, and the dashed lines are the two
  realizations on which the open tails are closed. \emph{Panel (b):} the reported histogram
  averaged over every forecaster-round, on the grid as the panel builder fixes it---historic open
  tails pooled, probabilities renormalized row by row within the grid regime, each bin drawn at
  the midpoint of its closed interval---with the revised grid superimposed. \emph{Panel (c):} the
  realized series binned on that same grid, against the average reported density.}
  \label{fig:grid}
  \small
  \textit{Note:} ECB-SPF one-year-ahead inflation densities, 1999Q1--2026Q2: $4{,}371$
  forecaster-rounds over $103$ rounds on the grid in force through 2024Q3, and $267$ over $8$
  rounds on the revised grid. Realized inflation is the year-on-year change in the euro-area HICP;
  panel (c) bins it at the $104$ target quarters realized to date. The average density places
  $0.9\,\%$ of its mass in the top bin against the $6.7\,\%$ of quarters realized there; inside
  the 2021Q3--2023Q4 rounds the top bin carries $8.5\,\%$ of reported mass on average, and
  $34.9\,\%$ in the round of 2022Q4. Script: \texttt{fig01\_histogram\_grid.py}.
\end{figure}

Figure~\ref{fig:grid} sets the grid beside the series it has to describe. The two supports agree
where it matters: realized year-on-year inflation has run between $-0.62\,\%$ and $10.62\,\%$ over
the sample, and $93\,\%$ of the quarters the survey forecasts fall inside the interior range on
which forecasters are asked to place mass, $[-1,5]$, the remainder in the top bin alone. Even where
the survey tracks the first moment poorly---through 2021--23 the consensus peaked at $5.1\,\%$
against a realization of $10.6\,\%$---the reported densities are therefore defined over the region
in which inflation has actually moved, so that their second and third moments can be read as
information about the dispersion and the asymmetry of beliefs rather than as an artifact of where
the questionnaire stops.

\subsection{Moments of a reported density}
\label{sec:moments}

\paragraph{Mean, median and mode.}
The mean, median and mode of each forecaster's subjective probability distribution (SPD) are read off the reported bins. Because the densities are reported as probability mass over bins, the procedures are the following.

\textbf{Mean ($\mu$)}: The mean is calculated as the weighted sum of the midpoints of the probability bins:

\begin{enumerate}
    \item Define the bin edges as intervals, $\text{bin\_edges} = [(a_1, b_1), (a_2, b_2), \ldots, (a_n, b_n)]$ where $(a_i, b_i)$ represents the lower and upper bounds of each bin.
    
    \item Calculate the midpoint for each bin $\text{midpoint}_i = \frac{a_i + b_i}{2}$
    \item Normalize the probabilities for each bin to ensure they sum to 100
    
    \item Compute the weighted mean $\mu = \sum_{i=1}^{n} \text{midpoint}_i \times P_i$ where $P_i$ represents the normalized probability associated with bin $i$.
\end{enumerate}

\textbf{Median}: The median is the value that divides the distribution into two equal parts, where half of the cumulative probability lies below it and half lies above. 
\begin{enumerate}
    \item Calculate the cumulative probabilities for each bin $\text{cumulative\_probs}_i = \sum_{j=1}^{i} P_j$ where $P_j$ is the probability for bin $j$.
    
    \item Identify the bin in which the cumulative probability first equals or exceeds 0.5 (the median percentile). Let the lower and upper edges of this bin be $a_i$ and $b_i$, respectively.
    
    \item Interpolate within the bin to find the exact median value $m$, using $$m = a_i + \left( \frac{0.5 - \text{cumulative\_probs}_{i-1}}{\text{cumulative\_probs}_i - \text{cumulative\_probs}_{i-1}} \right) \times (b_i - a_i)$$ where $\text{cumulative\_probs}_{i-1}$ represents the cumulative probability of the previous bin (or zero if $i = 1$).
\end{enumerate}

\textbf{Mode}: The mode is the value that occurs most frequently in the distribution, corresponding to the peak of the probability density function. 
\begin{enumerate}
    \item Identify the bins with the highest probability $\text{max\_prob} = \max(P_i)$ where $P_i$ represents the probability for bin $i$. All bins where $P_i = \text{max\_prob}$ are considered to have the highest probability.

    \item Determine the first and last bin indices, $i_\text{first}$ and $i_\text{last}$, that have this maximum probability.

    \item The bins carrying the maximum probability must be adjacent; otherwise the mode is not determined.

    \item Let $x$ be the lower edge of the first bin with the maximum probability and $\Delta$ the total width of the bins carrying it, $x = \text{bin\_edges}_{i_\text{first}, 0}, \quad \Delta = \text{bin\_edges}_{i_\text{last}, 1} - \text{bin\_edges}_{i_\text{first}, 0}$, and let
    \[
    c \;=\; x + \tfrac{1}{2}\Delta
    \]
    be their centre. One bin carrying the maximum puts $c$ at that bin's midpoint; an even number of adjacent bins sharing it puts $c$ at the edge where the two middle ones meet; an odd number puts $c$ at the midpoint of the middle one.%
\footnote{The widths are those of the closed bins of Section~\ref{sec:grid}. The bottom bin is degenerate, so it is a point mass lying wholly on one side of $c$; if it is itself the modal bin then $\Delta = 0$ and the mode is that point. The top bin is wide, so a modal block reaching into it has its centre correspondingly far to the right.}

    \item Compute the cumulative probabilities to the left ($l$) and to the right ($r$) of that centre,
    \[
    l \;=\; \sum_{j=1}^{n} \omega_j \, P_j, \qquad
    r \;=\; \sum_{j=1}^{n} (1-\omega_j) \, P_j, \qquad
    \omega_j \;=\; \min\left\{1, \; \max\left\{0, \; \frac{c-a_j}{b_j-a_j}\right\}\right\},
    \]
    where $n$ is the total number of bins, $a_j$ and $b_j$ are the edges of bin $j$, and $\omega_j$ is the share of bin $j$ lying below $c$, the mass being spread uniformly inside a bin exactly as it is for the quartiles. A bin entirely below the centre enters $l$ in full, a bin entirely above it enters $r$ in full, and the bin that contains the centre is split between the two in proportion to where $c$ falls inside it. Equivalently, $l$ is the interpolated cumulative distribution evaluated at $c$ and $r$ is its complement, so $l+r$ is the whole of the probability mass and $r/(r+l)$ is the share of it lying above the centre.

    \item Estimate the mode using the weighted formula:
    \[
    \text{mode} = x + \left(\frac{r}{r + l}\right) \Delta
    \]
    The mode is therefore a point of the modal bins, placed by the mass on either side of their centre: the peak sits at the centre when the two sides are equally heavy, and moves towards the heavier side otherwise, by $\tfrac{1}{2}\Delta\,(r-l)/(r+l)$. Measuring $l$ and $r$ about the centre rather than outside the modal bins also keeps the weight defined in every case---under the earlier convention a forecaster placing all the mass on the modal bins left $l = r = 0$ and the ratio undefined---and lets the whole reported distribution, not only the part outside the peak, locate the mode.
\end{enumerate}

Two cases fix the convention. With reported probabilities $0, 50, 50, 0$ the maximum is shared by the second and third bins and $c$ is the edge where they meet; each side of that edge carries half of the mass, $l = r = 50$, and the mode is the centre itself. With $10, 30, 30, 30, 0$ the maximum is shared by three bins and $c$ is the midpoint of the middle one; the two bins below it carry $40$ and the lower half of the middle bin carries $15$, so $l = 55$, while the upper half and the bin above give $r = 45$, and the mode sits just to the left of the centre, at $x + 0.45\,\Delta$.

\begin{center}
\begin{tikzpicture}[x=0.92cm, y=0.042cm, font=\footnotesize]
  \draw[dashed, thick] (2,-20) -- (2,58);
  \fill[black!45] (1,0) rectangle (3,50);
  \draw[black!85] (1,0) rectangle (2,50);
  \draw[black!85] (2,0) rectangle (3,50);
  \draw[black!60] (0,0) -- (4,0);
  \foreach \i in {0,1,2,3,4} \draw[black!60] (\i,0) -- ++(0,-0.09cm);
  \node at (0.5,10) {$0$};
  \node[white] at (1.5,25) {$50$};
  \node[white] at (2.5,25) {$50$};
  \node at (3.5,10) {$0$};
  \draw[<->] (1,64) -- (3,64) node[midway, above, inner sep=1.5pt] {$\Delta$};
  \node[below, inner sep=2pt] at (1,-2) {$x$};
  \fill (2,0) -- ++(-0.10cm,-0.26cm) -- ++(0.20cm,0) -- cycle;
  \node[fill=white, inner sep=1pt] at (2,-16) {$c$};
  \draw[<->] (0,-34) -- (2,-34) node[midway, above, inner sep=1.5pt] {$l=50$};
  \draw[<->] (2,-34) -- (4,-34) node[midway, above, inner sep=1.5pt] {$r=50$};
  \node at (2,-56) {mode $= c$};
  \node at (2,-80) {(a) $0,\,50,\,50,\,0$};
  \begin{scope}[xshift=5.2cm]
    \draw[dashed, thick] (2.5,-20) -- (2.5,38);
    \fill[black!18] (0,0) rectangle (1,10);
    \fill[black!45] (1,0) rectangle (4,30);
    \draw[black!85] (0,0) rectangle (1,10);
    \draw[black!85] (1,0) rectangle (2,30);
    \draw[black!85] (2,0) rectangle (3,30);
    \draw[black!85] (3,0) rectangle (4,30);
    \draw[black!60] (0,0) -- (5,0);
    \foreach \i in {0,1,2,3,4,5} \draw[black!60] (\i,0) -- ++(0,-0.09cm);
    \node at (0.5,18) {$10$};
    \node[white] at (1.5,22) {$30$};
    \node[white, fill=black!45, inner sep=1pt] at (2.5,22) {$30$};
    \node[white] at (3.5,22) {$30$};
    \node at (4.5,10) {$0$};
    \node[white] at (2.22,9) {$15$};
    \node[white] at (2.78,9) {$15$};
    \draw[<->] (1,44) -- (4,44) node[midway, above, inner sep=1.5pt] {$\Delta$};
    \node[below, inner sep=2pt] at (1,-2) {$x$};
    \fill (2.35,0) -- ++(-0.10cm,-0.26cm) -- ++(0.20cm,0) -- cycle;
    \node[fill=white, inner sep=1pt] at (2.5,-16) {$c$};
    \draw[<->] (0,-34) -- (2.5,-34) node[midway, above, inner sep=1.5pt] {$l=55$};
    \draw[<->] (2.5,-34) -- (5,-34) node[midway, above, inner sep=1.5pt] {$r=45$};
    \node at (2.5,-56) {mode $= x+0.45\,\Delta$};
    \node at (2.5,-80) {(b) $10,\,30,\,30,\,30,\,0$};
  \end{scope}
\end{tikzpicture}
\end{center}

\noindent In both panels the shaded bins are the ones carrying the maximum probability, the dashed line is their centre $c$, and $l$ and $r$ are the mass to its left and to its right, the bin that contains $c$ contributing to each in proportion. In panel~(b) that bin is split in half, $15$ each side, which is what moves $l$ to $55$ and $r$ to $45$ and places the mode left of the centre.

\paragraph{Variance.}
The variance is the probability-weighted average of the squared deviations of the bin
midpoints from the mean, $\operatorname{Var} = \sum_{i=1}^{n} (\text{midpoint}_i - \mu)^2
\, P_i$, and the standard deviation is its square root. The coefficient of variation,
$\sqrt{\operatorname{Var}}/\mu$, is not used: it diverges as the mean approaches zero,
which is exactly the region the euro-area sample spends a decade in. The normalization of
Section~\ref{sec:nu_envelope} is the target-centred alternative.

\paragraph{Quartiles and asymmetry.}
Bowley coefficient of skewness is a measure of asymmetry in a distribution, based on the quartiles \(Q_1\) (25th percentile), \(Q_2\) (median or 50th percentile), and \(Q_3\) (75th percentile). It is calculated using the formula:
\begin{equation}
S = \frac{(Q_3-Q_2)-(Q_2-Q_1)}{Q_3-Q_1}
\end{equation}
This coefficient ranges from -1 to 1, where negative values indicate left skewness (a distribution with a long tail to the left), and positive values indicate right skewness.

To illustrate, consider the following inflation forecast bins with their corresponding probabilities:

\begin{tabular}{|c|c|}
\hline
\textbf{Inflation Bins} & \textbf{Probabilities} \\
\hline
$[-1.5, -1]$ & 10 \\
\hline
$[-1, -0.5]$ & 25 \\
\hline
$[-0.5, 0]$ & 35 \\
\hline
$[0, 0.5]$ & 25 \\
\hline
$[0.5, 1]$ & 5 \\
\hline
\end{tabular}

$Q_1$ (25th percentile) lies within the bin $[-1, -0.5]$ and can be interpolated as $Q_1 = -1 + \frac{15}{25} (0.5) = -0.7$

In extreme situations---such as 2023, when some forecasters placed more than $25\%$ of the probability mass in the bin $[5, +\infty[$---the standard interpolation breaks down. Such cases are treated as conveying a clear signal of positive skewness. Symmetrically, when more than $25\%$ is placed in the bin $]-\infty, 1]$, it is treated as a clear signal of negative skewness, and the historical quartiles of the skewness coefficient are applied.

\begin{remark}[Sensitivity to the skewness measure]
\label{rem:ac_skewness_sensitivity}
The AC construction requires a bounded asymmetry measure $A_{i,t}\in[-1,1]$, a property satisfied by Bowley's skewness $B=(Q_3+Q_1-2Q_2)/(Q_3-Q_1)$.
The Pearson mode skewness $(\text{mean}-\text{mode})/\sigma$ and the classical moment-based skewness $m_3/m_2^{3/2}$ are unbounded and more sensitive to outliers in sparse histogram bins.
Replacing Bowley's skewness with either alternative in the AC formula yields qualitatively identical time-series patterns: the quarterly AC series constructed with Bowley's formula, Pearson mode, and moment-based skewness exhibit pairwise correlations exceeding~0.95.
The quantitative results in this paper are therefore robust to the choice of skewness estimator; Bowley's measure is retained for its natural boundedness and computational simplicity.
\end{remark}

\subsection{Individual densities, not the average}
\label{sec:individual}

Two objects can be called the uncertainty of a survey round: the average of the
individual variances, and the variance of the density obtained by averaging the
histograms. They are not the same, and the difference is disagreement.

\begin{figure}[tbp]
  \centering
  \includegraphics[width=\linewidth]{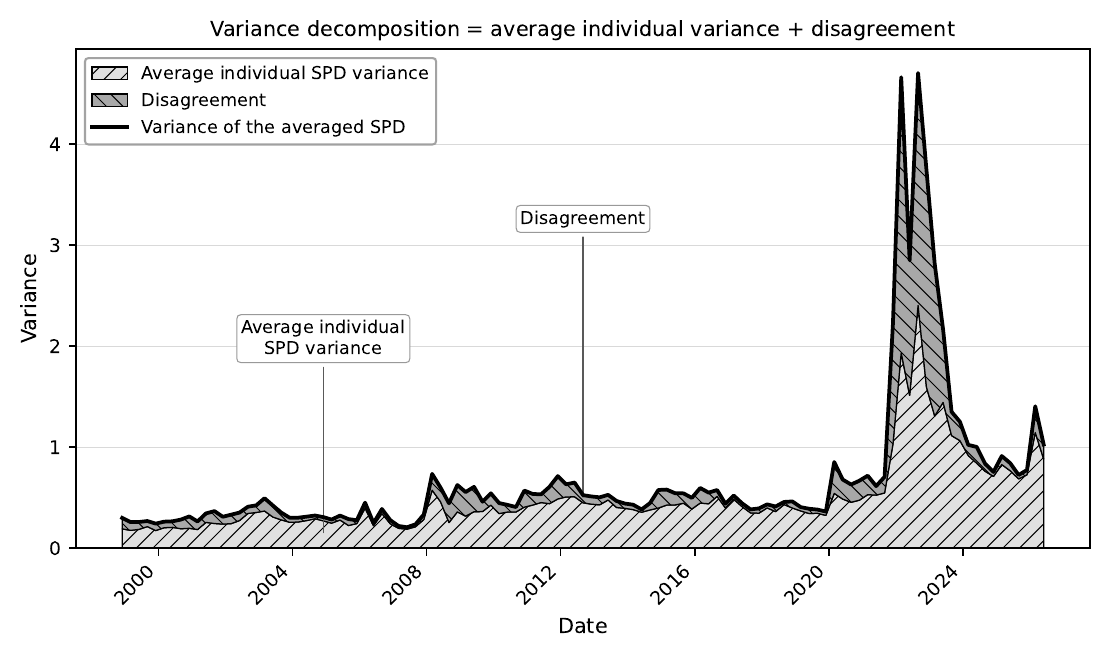}
  \caption{Variance decomposition of inflation forecast uncertainty. The lower band, hatched to the
  right and lightly shaded, is the cross-sectional average of the individual SPD variances; the upper
  band, hatched to the left and darker, is disagreement (constructed to satisfy
  identity~\eqref{eq:var-mixture}); the solid black line is the variance of the averaged SPD, which the
  two bands add up to by construction. Both bands are also labelled inside the panel, so the figure
  reads in black and white. Script: \texttt{fig02\_variance\_decomposition.py}.}
  \label{fig:var_decomposition}
\end{figure}

Formally, let $\mu_i$ denote the mean of forecaster $i$'s distribution and $\sigma_i^2$ their subjective variance. The variance of the mixture distribution satisfies
\begin{equation}\label{eq:var-mixture}
\mathrm{Var}\!\left(\tfrac{1}{N}\sum_{i=1}^N X_i\right)
= \underbrace{\tfrac{1}{N}\sum_{i=1}^N \sigma_i^2}_{\text{average individual uncertainty}}
+ \underbrace{\tfrac{1}{N}\sum_{i=1}^N \left(\mu_i - \bar{\mu}\right)^2}_{\text{disagreement}},
\end{equation}
where $\bar{\mu} = \tfrac{1}{N}\sum_{i=1}^N \mu_i$ is the average forecast mean. This decomposition follows \citet{zarnowitz1987consensus} and is illustrated in Figure \ref{fig:var_decomposition} for inflation forecasts.

\paragraph{Remark.}
The identity in \eqref{eq:var-mixture} holds exactly when the individual SPDs are defined over the same, mutually exclusive and collectively exhaustive partition of outcomes.
If this condition is violated---e.g.\ if respondents' bins do not align perfectly---then the averaged SPD will mechanically exhibit excess dispersion.
In such cases, the variance of the averaged histogram is artificially inflated relative to the theoretical mixture variance, and the residual overstates true disagreement.
This highlights that this decomposition is only valid under the assumption of a common support and consistent binning across individual SPDs.
In the ECB-SPF every respondent answers on the same grid, and Figure~\ref{fig:var_decomposition} is built on that identity: each density is renormalized with a missing cell read as no mass, so the two bands add up to the variance of the averaged density exactly, and disagreement is the population variance of the individual means and nothing else.

\begin{figure}[tbp]
    \centering
    \includegraphics[width=0.6\linewidth]{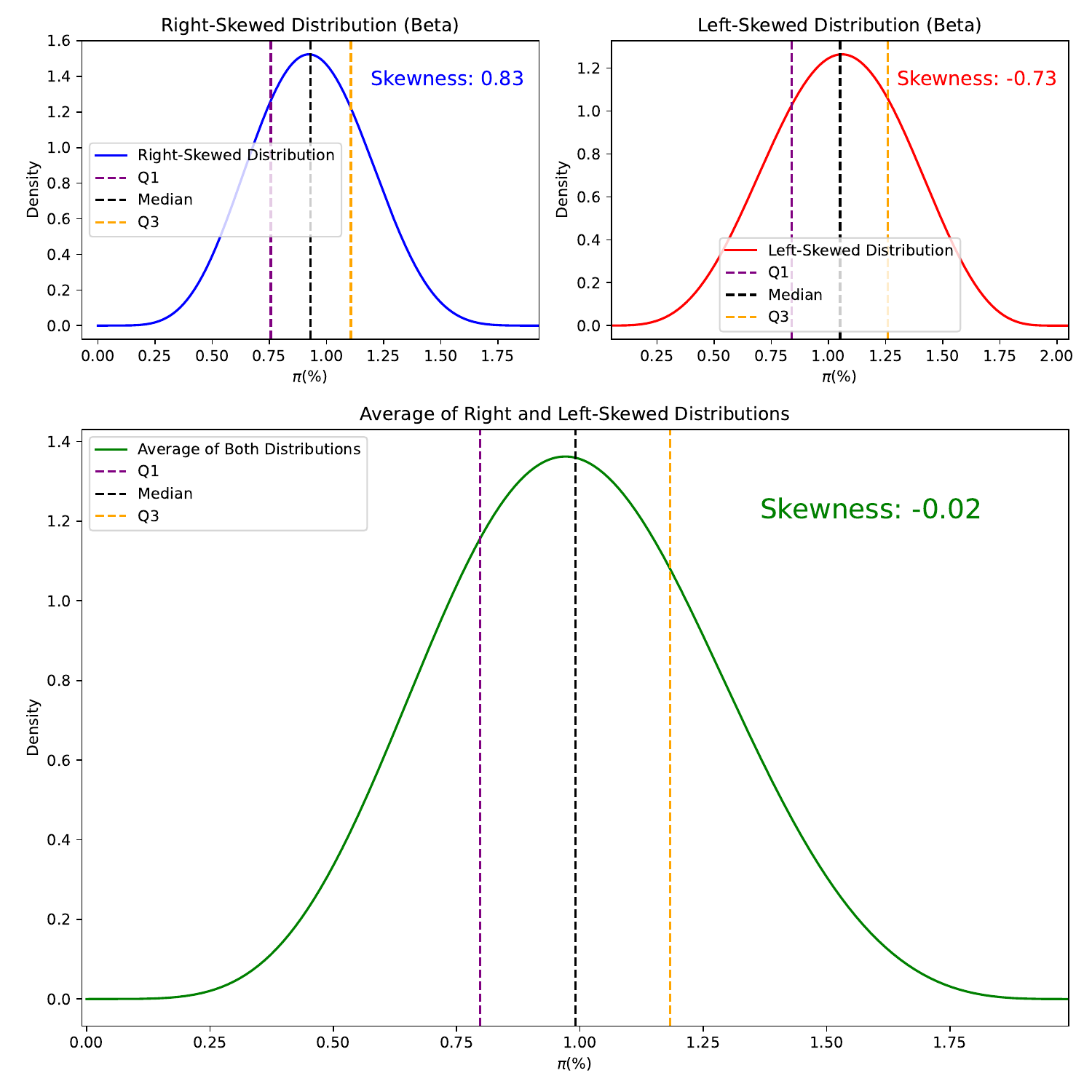}
    \caption{Illustration of Skewness Variation in Individual vs. Average SPD}
    \label{fig:two_SPD_skewness_illustration}
    \small
\textit{Note:} The illustration is constructed from two scaled beta distributions, with parameters $(a,b)=(5.5,5)$ to generate a mildly left-skewed shape and $(a,b)=(7,8)$ to generate a more right-skewed shape, both rescaled by a factor of $2$. Quartiles and Bowley's skewness are calculated for each distribution and their simple average, illustrating how aggregation can obscure the underlying asymmetries of individual distributions. Script: \texttt{fig03\_two\_densities.py}.
\end{figure}

The third moment behaves worse under averaging, because aggregation can reverse its sign.
Following \citet{kotz2012laplace}, Bowley's skewness of a sum of independent random variables can carry a counterintuitive sign. This favors using the average of individual skewness measures rather than the skewness of the average random variable. Figure~\ref{fig:two_SPD_skewness_illustration} shows the
mechanism: two forecasters may both report right-skewed densities and, because they
disagree about where the centre lies, average to a left-skewed histogram. The
construction is standard for the asymmetric Laplace law.

Both facts point the same way, and fix a rule used everywhere below: every statistic in
this paper --- variance, standard deviation, quartiles, asymmetry --- is computed from a
single forecaster's reported density, and only the resulting statistics are averaged
within a survey round. Nothing is computed from an averaged histogram.
    \section{Normalized Uncertainty}
\label{sec:nu}

Do the macroeconomic effects commonly attributed to inflation uncertainty reflect genuine belief imprecision, or do they largely proxy for high-inflation regimes?
If dispersion increases predictably when expectations deviate from a policy anchor, then raw uncertainty measures conflate two conceptually distinct objects: (i) a \emph{structural component}---the predictable widening of the forecast distribution when the expected level drifts from target---and (ii) \emph{genuine uncertainty}---fluctuations in belief precision that are orthogonal to the level of expectations.
Disentangling these two components is critical.
Without doing so, one cannot determine whether it is inflation uncertainty \textit{per se} that weakens monetary policy transmission, de-anchors expectations, and depresses growth, or whether these effects merely reflect the level of inflation itself.
This contamination problem is not specific to inflation or surveys---it arises whenever variance is used as a state variable in environments where first-moment movements mechanically widen distributions---but it is particularly consequential in the inflation context, where decades of empirical work rest on potentially confounded uncertainty measures.

The construction proceeds in two steps. Section~\ref{sec:nu_envelope} writes the measured
variance as an envelope in the distance from target times a residual, shows why dividing by
the envelope stabilizes it --- that division is the measure --- fits the envelope on the
euro-area panel, states the two operational series it yields and reports what the division
removes. Why the dispersion
of a reported density moves with the distance from target, and where that regularity sits in
the literature, is Section~\ref{sec:distance_volatility}.

\subsection{The envelope and its estimates}
\label{sec:nu_envelope}

Write the measured variance of a reported density as a structural component evaluated at
the observed distance, times a residual:
\begin{equation}
  \label{eq:envelope}
  \operatorname{Var}_t \;=\; V_{\mathrm{struct}}(d_t)\times NU^2_t,
  \qquad
  d_t \;\equiv\; \mu_t-\mu^{*},
\end{equation}
where the distance carries its sign and $\mu^{*}=2\,\%$ is the announced target. The form of
the envelope $V_{\mathrm{struct}}$ is specified below; what comes first needs only that it be
positive and evaluated at the observed distance.

\paragraph{Stabilizing the variance.}
The repair for dispersion that moves with an observable state is classical: the
variance-stabilizing transformation (VST) \citep{anscombe1948transformation,
nelder1972generalized}, of the same family as the Box--Cox power transformation
\citep{box1964analysis}. If the variance of a random variable~$X$ depends on a state through a
known function $h$---here $h(d)=V_{\mathrm{struct}}(d)$, the envelope
of~\eqref{eq:envelope}---then dispersions read at different distances are readings on
rulers of different lengths, and the delta method supplies the repair: any transformation
with $f'\propto 1/\sqrt{h}$ makes $\operatorname{Var}(f(X))$ state-free to first order.
The form of $h$ picks the family. A variance linear in the distance is the
compound-Poisson signature of the mechanism the companion paper derives---an arrival
rate times a quantum---and for that mean--variance shape the stabilizer is the square
root, for the same reason that $\sqrt{X}$ stabilizes a Poisson count
\citep{anscombe1948transformation}; the kink at the target that the envelope turns out to
have poses no obstacle, the envelope staying continuous and positive through it. Because the distance is observed,
the correction need not stop at first order: dividing the measured variance by the
envelope \emph{at the observed distance} removes the dependence exactly under the
multiplicative reading of~\eqref{eq:envelope}---no expansion, no retransformation---and
that form is the measure. \emph{Normalized Uncertainty} (NU) is the measured standard
deviation in units of the standard deviation the envelope predicts at the observed distance,
\begin{equation}
  \label{eq:NU_definition}
  \mathrm{NU}_t
    \;\equiv\;
    \frac{\sqrt{\operatorname{Var}_t}}{\sqrt{V_{\mathrm{struct}}(d_t)}},
\end{equation}
a square root of a ratio because stabilization lives on the standard-deviation scale.%
\footnote{A simple coefficient of variation $\sqrt{\operatorname{Var}(X_t)}/|\mu_t|$ is not suitable: inflation values close to zero would cause the measure to diverge. The target-centered VST remains robust to both the dependence on the mean and the near-zero inflation problem, and is equally applicable to economic growth rates where the anchor is potential output.} In a word, $\mathrm{NU}$ is
\emph{residual} uncertainty: the part of measured dispersion the anchor does not
explain.

\paragraph{The form of the envelope.}
The division needs the envelope itself. It is specified in the signed distance, with a
separate slope on each side of the target,
\begin{equation}
  \label{eq:vstruct}
  V_{\mathrm{struct}}(d) \;=\; a + b_-\,(-d)_+ + b_+\,(d)_+ ,
  \qquad (x)_+\equiv\max(x,0).
\end{equation}

\begin{figure}[tbp]
  \centering
  \includegraphics[width=\linewidth]{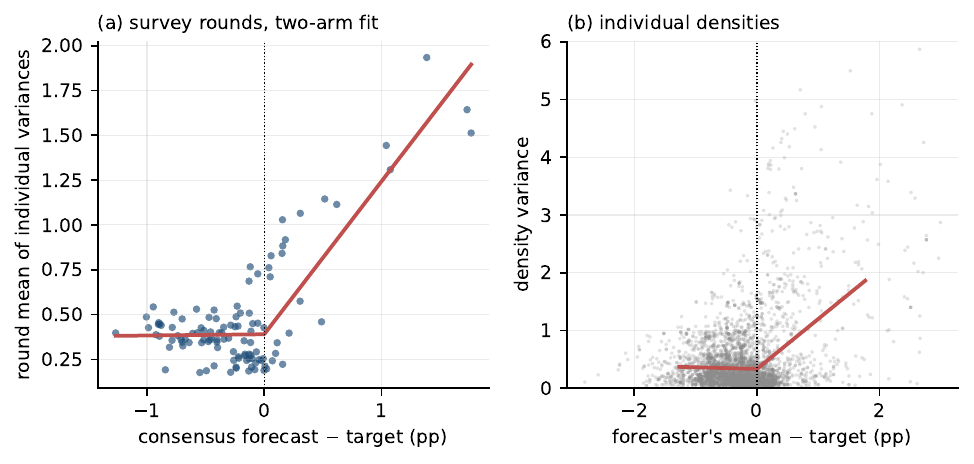}
  \caption{\textbf{The variance--distance envelope on the euro-area panel.} Round-mean variance of the
  individual one-year-ahead inflation densities against the signed distance of the consensus
  forecast from the $2\,\%$ target, with the two-arm fit of~\eqref{eq:vstruct}; the
  individual-level scatter beside it. Certified panel; the round panel excludes the one round
  whose consensus lies outside $[-1,5]$ (2022Q4), the individual panel the densities whose
  mean does. Script: \texttt{fig04\_arms\_fit.py}.}
  \label{fig:arms_fit}
\end{figure}

\paragraph{Estimates.}
Table~\ref{tab:arms} reports the envelope on the certified euro-area panel: individual
one-year-ahead inflation densities aggregated to the survey round, rounds whose consensus lies
outside $[-1,5]$ excluded (Section~\ref{sec:grid}), $n=109$ rounds through 2026Q2 and $110$
through the latest round, with Newey--West standard errors at four lags. What the measure
divides is a forecaster's own density, so the envelope that matters is that of the round mean
of the individual variances, $W_t$. Through 2026Q2 it is $\hat a=0.383$, $\hat b_-=+0.004$
($t=0.05$) and $\hat b_+=0.855$ ($t=12.22$), with $R^2=0.705$ --- the fit the companion paper
reports as its survey law; the consensus forecast ranges over $[0.73,3.76]$, so both arms are
identified by the data and not by extrapolation. Extending the window to the latest round
moves nothing that matters: $\hat a=0.390$, $\hat b_-=-0.007$, $\hat b_+=0.853$, $R^2=0.697$.
On those rounds two Wald tests read the shape directly: $b_-=0$ is not rejected
($\chi^2_1=0.005$, $p=0.94$), and $b_-=b_+$ --- the symmetric form --- is rejected
($\chi^2_1=91.9$, $p<0.001$). The disagreement between forecasters has the same
shape with a steeper upper arm and a measurable lower one --- $1.028$ above ($t=8.93$) and
$0.113$ below ($t=2.36$) through 2026Q2 --- and the total variance, $W_t$ plus that
disagreement, inherits the individual shape, $W_t$ carrying $73\,\%$ of it. At the level of
the individual density, pooled over forecaster-rounds, the upper arm is $0.864$.
Figure~\ref{fig:arms_fit} draws the fit.

\paragraph{One ratio for calibration.}
Only the ratio $r_+\equiv b_+/a$ enters the measure. Dividing numerator and denominator of
the normalization by $\sqrt{a}$ rescales every value of the series by the same constant and
changes nothing that is read from it --- not its dynamics, not the ranking of two dates, not
its correlation with anything. On the panel $r_+=0.855/0.383=2.23$.

Only $r_+=b_+/a$ has to be chosen, and there are two defensible choices. They are reported
side by side throughout, because each answers a different requirement.

The \emph{fitted} series takes the ratio from the envelope of $W_t$ fitted through 2026Q2,
\begin{equation}
  \label{eq:nu_fit}
  \mathrm{NU}^{\mathrm{fit}}_{i,t}
  = \frac{\sigma_{i,t}}{\sqrt{1+r_+\,(\mu_{i,t}-\mu^{*})_+}},
  \qquad r_+=2.23 ,
\end{equation}
with $\sigma_{i,t}$ forecaster $i$'s density standard deviation in round $t$ and the round
series the mean of the individual values. It is the series the companion paper's
regressions use, and it is the right choice when the envelope is the object of interest.
Its cost is that it moves: re-estimate the envelope on a longer sample and every historical
value of the series changes, which is a look-ahead the measure does not need.

Table~\ref{tab:arms} makes that cost concrete. Fitted on the rounds before 2020 the same
envelope returns $r_+=1.50$, with a delta-method standard error of $0.63$, against $2.23$
through 2026Q2. The reason is the support the sample happens to contain: before 2020 the
consensus gap ran over $[-1.27,+0.49]$ and nine rounds of eighty-four sat above target, so the
upper arm is identified by almost nothing, and what identifies it afterwards is the 2021--23
overshoot. The two periods do not share one envelope ($\chi^2_3=380$, $p<0.001$), although the
difference between the two ratios on its own is within one standard error of zero --- which is
the point rather than an objection: a fitted ratio is a dated object, it inherits the range of
distances its window contains, and it will move again when the next episode widens that range.
The calibrated series is immune to this by construction.

The \emph{calibrated} series sets the ratio to one,
\begin{equation}
  \label{eq:nu_cal}
  \mathrm{NU}^{\mathrm{cal}}_{i,t}
  = \frac{\sigma_{i,t}}{\sqrt{1+(\mu_{i,t}-\mu^{*})_+}} .
\end{equation}
Nothing in it is estimated, so it is reproducible from the raw histograms alone, fixed once
and for all, and portable to a survey, a horizon or a country where the envelope has not
been fitted --- which is also why the growth measure of Section~\ref{sec:growth} is built
this way. It is the series the companion paper draws and the one its agreement check with
an independent text-based index is computed on. Its cost is that the denominator understates
the envelope the euro-area data actually show, so it removes less of the level component
than the fitted series does.

The two series correlate at $0.95$ over the full sample. Section~\ref{sec:simulation}
reports what misspecifying the ratio costs in a controlled setting; the short answer is
that the ranking of the corrected and the raw measure survives a misspecification of $r_+$
by a factor of several, which is what licenses publishing both.

\begin{remark}[Portability beyond survey densities]
\label{rem:portability}
The variance--distance framework derived above is not specific to inflation surveys.
The affine relationship requires only two structural ingredients: (i)~a reference point~$\mu^*$ that anchors expectations, and (ii)~economic mechanisms---that generate level-dependent variance.
These ingredients are present in many empirical settings.
For GDP growth forecasts, the natural anchor is potential output growth, and the resulting \emph{Normalized Growth Uncertainty} (NGU) is deployed in Section~\ref{sec:growth}.
For option-implied densities, the forward price provides the reference point, and implied volatility exhibits well-documented dependence on the distance of the underlying from key levels.
For firm-level expectations on sales, prices, or investment, industry-specific benchmarks or historical means can serve as anchors.
More broadly, the NU normalization applies to any predictive distribution in which raw dispersion mechanically co-moves with first-moment deviations from a benchmark---a pervasive feature of economic forecasting environments---and the variance-stabilizing transformation of Section~\ref{sec:nu_envelope} is directly portable.
This portability is not merely conceptual: NU already enters as a regressor in related work on monetary policy and credit supply \citep{vansteenberghe2025monetary}.
\end{remark}

\begin{figure}[tbp]
  \centering
  \includegraphics[width=0.95\linewidth]{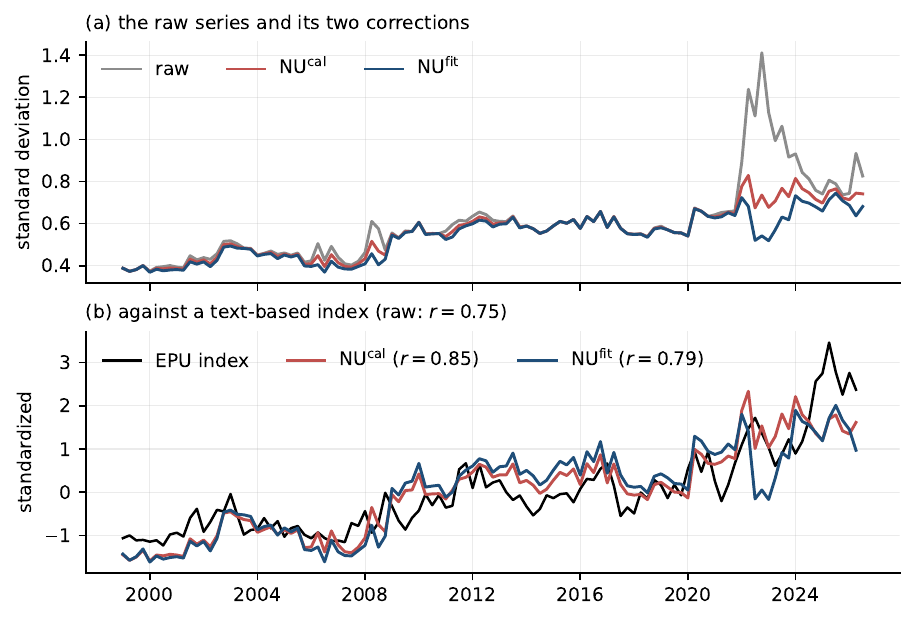}
  \caption{\textbf{The correction in the euro-area series.} \emph{Panel (a):} raw survey uncertainty
  and its two corrected counterparts, $\mathrm{NU}^{\mathrm{fit}}$ and
  $\mathrm{NU}^{\mathrm{cal}}$, quarterly. \emph{Panel (b):} the corrected series against the
  euro-area economic policy uncertainty index of \citet{baker2016measuring}, both standardized
  over their overlap. Script: \texttt{fig05\_nu\_series.py}.}
  \label{fig:nu_series}
\end{figure}

\paragraph{The series, and an outside check.}
Figure~\ref{fig:nu_series} draws the raw round-mean standard deviation against the two
corrected series. The three agree through the anchored decades, where the consensus forecast
sits at the target and the denominator is flat, and separate exactly where the distance
opens: the 2021--2023 surge, and again in 2026.

Whether the separation is an improvement cannot be settled inside the survey, so it is put
to an outside instrument: the euro-area economic policy uncertainty index of
\citet{baker2016measuring}, built from newspaper text, which shares none of the survey's
construction. Over the 110 quarters the two overlap, the correlation is $0.746$ for the raw
series, $0.848$ for the calibrated series and $0.785$ for the fitted one. Both corrections
therefore buy agreement with an independent proxy.

\paragraph{How much of the separation is a measurement.}
The height of the gap between the raw and the corrected series in 2022--2023 should not be
read as a quantity. Both series descend from a histogram whose top bin is open-ended, and
in those rounds enough probability mass sits inside that bin that the raw second moment
partly reflects where the tail is closed rather than what forecasters believe
(Section~\ref{sec:grid}). The normalization damps the problem, since the denominator is
largest exactly where the grid fails, but it does not remove it. That the two series
separate is a fact about them; the size of the separation in those quarters is not a
measurement.

\medskip
The second moment is one half of what a reported density carries. The next section turns
to the third.
    \section{Asymmetry Coherence}
\label{sec:ac}

Section~\ref{sec:nu} exploited that the conditional variance of subjective forecast distributions increases with the distance of expectations from the policy anchor, and developed the Normalized Uncertainty (NU) measure to isolate genuine belief imprecision from this structural dependence. Yet variance captures only the \emph{width} of the distribution, not its \emph{shape}. A natural complementary question arises: does the asymmetry of subjective distributions also vary systematically with the deviation of expectations from anchor? If so, the third moment of forecast distributions encodes a structural signal about directional risk---about which side the risks are on, and how coherently---that is conceptually distinct from, and only partly overlapping with, the uncertainty signal captured by NU. In monetary policy communication, financial stability assessments, and risk management, the concept of the ``balance of risks'' is ubiquitous. Yet there exists no standard statistical operationalization of this concept. This section provides one.

\begin{figure}[tbp]
  \centering
  \includegraphics[width=\linewidth]{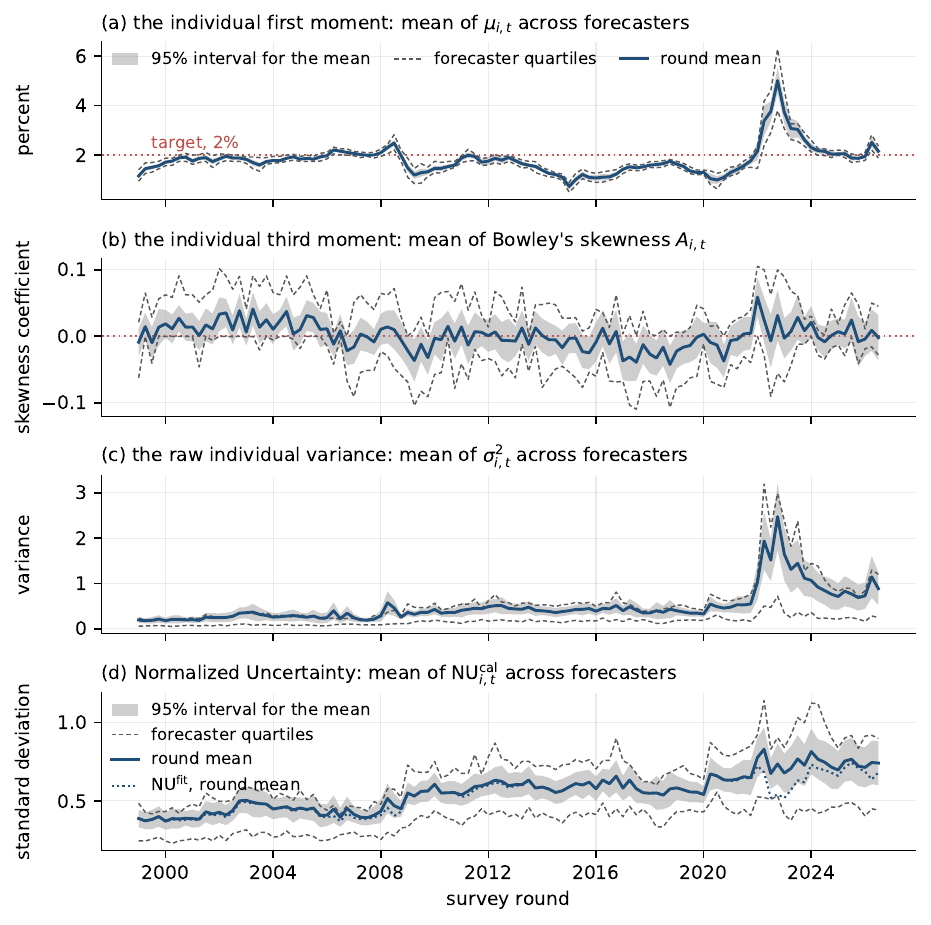}
  \caption{\textbf{The moments of the reported densities, round by round.} Cross-forecaster mean of
  \emph{(a)} the individual density means $\mu_{i,t}$, \emph{(b)} the individual Bowley skewness
  $A_{i,t}$, \emph{(c)} the raw individual variance $\sigma^2_{i,t}$---the object $W_t$ the envelope of
  Section~\ref{sec:nu_envelope} is fitted on---and \emph{(d)} Normalized Uncertainty in its calibrated
  form, with the fitted series dotted beside it. The shaded band is the ninety-five per cent interval for
  the round mean, $\bar{x}\pm t_{0.975,\,N_t-1}\,s/\sqrt{N_t}$, and the dashed lines are the quartiles of
  the forecasters reporting that round; the dotted horizontal line is the target in~(a) and symmetry
  in~(b).}
  \label{fig:moments}
  \small
  \textit{Note:} ECB-SPF one-year-ahead inflation densities, 1999Q1--2026Q3: $4{,}624$ forecaster-rounds
  over $111$ rounds, $30$ to $63$ forecasters per round; densities with a degenerate variance are dropped.
  The interval of the first moment excludes the $2\,\%$ target in $96$ rounds, $77$ of them below it; that
  of the third moment excludes zero in $19$. Against the 2015--2019 average, the peak of the 2022--2023
  rounds is $6.1$ times the raw variance and $1.41$ times $\mathrm{NU}^{\mathrm{cal}}$. Script:
  \texttt{fig06\_moments\_intervals.py}.
\end{figure}

An empirical regularity motivates this investigation. In the ECB-SPF inflation data, the mean and the asymmetry of the reported densities are positively related (Table~\ref{tab:momentscorr}): across survey rounds the correlation between the average mean and the average Bowley skewness is $0.36$, and across individual forecaster-rounds it is $0.11$ --- positive at both levels, and an order of magnitude weaker at the level where the third moment is actually measured, which is the noise problem this section is built around: higher expected inflation levels are systematically associated with greater right-tail risk, and lower expectations with left-tail risk. This pattern suggests that forecasters' perceived asymmetry is not idiosyncratic noise but responds to the position of expectations relative to the inflation target. Figure~\ref{fig:moments} sets that regularity beside the moments this paper corrects, round by round: the cross-forecaster mean of the individual first moment, of Bowley's skewness, of the raw variance and of Normalized Uncertainty, each with the ninety-five per cent interval of the round mean and the quartiles of the cross-section. Two features of it run through what follows. The first moment is precisely measured and sits away from target in ninety-six of the hundred and eleven rounds, while the mean third moment cannot be told from symmetry in ninety-two of them: read on its own the average asymmetry of a round carries little, which is what the coherence weight of Section~\ref{sec:AC_measure} is for. And the 2021--23 episode multiplies the round-mean raw variance by six against its 2015--19 average while leaving the normalized series within half again of it---the correction of Section~\ref{sec:nu} working across forecasters as it does through time.

This section builds an \emph{Asymmetry Coherence} (AC) measure --- an operational
formalization of the ``balance of risks'' --- that extracts that directional signal.
Section~\ref{sec:distance_asymmetry} sets out why a third moment has to be read jointly
with the first, Section~\ref{sec:AC_measure} gives the construction, and
Section~\ref{sec:AC_empirical} reports the euro-area series and how far it overlaps with
Normalized Uncertainty. The measures of Section~\ref{sec:nu} and of this section are built
from the same reported densities, one round at a time, under the conventions of
Section~\ref{sec:data}.

\subsection{Why the third moment must be read with the first}
\label{sec:distance_asymmetry}

The macroeconomic and finance literatures increasingly use skewness as a proxy for directional risk, tail risk, or the balance of risks.
In asset pricing, \citet{kraus1976skewness} and \citet{harvey2000conditional} establish that skewness is a priced risk factor.
In macro, \citet{ball1995relative} link aggregate inflation to the skewness of relative price changes; \citet{kitsul2013economics} extract option-implied inflation densities with time-varying skewness; and \citet{bianchi2022belief} show that SPF inflation-skewness shifts predict subsequent inflation and output movements.
More recently, \citet{detaming} use skewness augmentations of UCSV models for density forecasting.

The common assumption across these applications is that observed skewness can be interpreted directly as directional risk.
The third moment, however, is often measured too noisily for such a reading to be warranted.
In survey data, subjective probability distributions are reported in sparse bins; in option-based settings, densities are recovered from a limited set of strikes and maturities; and in model-based approaches, skewness remains exposed to specification error.
In these environments, raw skewness or asymmetry should not be treated as fully informative.

In the ECB-SPF, the first moment is positively related to individual asymmetry, but it explains only a very small fraction of its variation.
In Table~\ref{tab:skew_dev_regs}, the signed deviation of the SPD mean from target enters significantly, yet the corresponding $R^2$ is about one per cent in the baseline specification.
So the issue is not that raw asymmetry is mechanically driven by the first moment, but that it is too noisy to be interpreted confidently on its own.
The AC measure addresses this problem by using the first moment as a disciplining device: it retains only the component of asymmetry that is coherent with the central forecast, yielding a more operational measure of directional risk and of the balance of risks.

This subsection carries the framework of Section~\ref{sec:distance_volatility} from the second moment to the third. Throughout this subsection, $\mu_3(X_{t+1}\mid\mathcal{I}_t) \equiv \mathbb{E}\bigl[(X_{t+1} - \mu_t)^3 \mid \mathcal{I}_t\bigr]$ denotes the conditional third central moment of the forecaster's subjective predictive distribution.

The envelope of Section~\ref{sec:nu_envelope} reads the distance through its two arms, so it records how far a forecast sits from the anchor and on which side of it; what a variance cannot record is which way the risk leans. The third moment, by contrast, is inherently directional: a positive $\mu_3$ indicates right-skew (upside risk), a negative $\mu_3$ left-skew (downside risk). The conditional third moment is therefore modeled as a function of the \emph{signed} deviation:
\begin{equation}
  \label{eq:skew_signed_dev}
  \mu_3(X_{t+1}\mid\mathcal{I}_t) \;=\; h(\mu_t - \mu^*),
\end{equation}
where $h\colon\mathbb{R}\to\mathbb{R}$ is differentiable with $h(0)=0$: at the anchor, symmetric distributions carry no directional signal. The slope $h'(0)$ determines whether skewness increases in the same direction as the deviation (\emph{coherent} asymmetry, $h'(0) > 0$) or in the opposite direction (\emph{anti-coherent} asymmetry, $h'(0) < 0$). The maintained assumptions below provide independent economic channels that determine the sign and magnitude of this slope.

The problem with raw skewness is not primarily that it is mechanically driven by the first moment.
Rather, the problem is interpretability.
In many empirical environments, the third moment is difficult to estimate with precision: survey densities are reported on sparse histogram bins, option-implied densities are recovered from a limited range of strikes and maturities, and model-based skewness remains sensitive to specification error.
As a result, observed asymmetry often combines a potentially meaningful directional component with substantial measurement noise.

A useful benchmark is the case of \emph{coherent asymmetry}, in which the sign of the third moment agrees with the sign of the deviation of the median forecast from the anchor.
Formally, one may write
\begin{equation}
  \label{eq:skew_distance_preview}
  \mu_3(X_{t+1}\mid\mathcal{I}_t)
  = c\,(Q_t-\mu^*) + \eta_t,
\end{equation}
where \(c>0\) and \(\eta_t\) collects measurement error, specification noise, and idiosyncratic shape variation unrelated to the directional signal.
When \(Q_t>Q^*\), the case \(c>0\) implies positive skewness: forecasters not only expect inflation above target, but also assign relatively more probability mass to upside outcomes.
From a policy perspective, this is an informative configuration.
It suggests that the expected policy response is not perceived as forceful enough to eliminate upside inflation risk.
Symmetrically, when \(Q_t<Q^*\), coherent negative skewness indicates perceived downside risk around an already weak central outlook.

Raw skewness alone cannot distinguish this economically meaningful case from a distribution whose estimated asymmetry is large but directionally uninformative.
A positive third moment is not necessarily a strong upside-risk signal unless it is read jointly with the location of the distribution.
This is why the relevant object is not asymmetry per se, but \emph{asymmetry coherence}: whether the third moment reinforces the message of the first moment or contradicts it.
When the first and third moments have the same sign relative to the anchor, the predictive density sends a \emph{strong} directional signal.
When they point in opposite directions, the signal is \emph{weak}: the center of the distribution and its tail configuration do not tell the same story.
The illustrative figure in Section~\ref{sec:AC_measure} makes this distinction visually by contrasting coherent and incoherent combinations of the first and third moments.

The AC index, developed in Section~\ref{sec:AC_measure}, operationalizes this idea by filtering raw asymmetry through sign agreement with the central forecast.
It therefore does not treat the third moment as fully informative on its own; instead, it extracts the component of asymmetry that is consistent with the directional message conveyed by the first moment.

\subsection{The construction}
\label{sec:AC_measure}

When asymmetry is coherent on average ($c>0$), the third moment of a reported density shares the sign of the deviation of its centre from the anchor. That is what makes a directional signal extractable from expectations data. However, two practical obstacles prevent the direct use of skewness as a policy indicator.

First, skewness estimates are inherently fragile. The third moment is highly sensitive to noise and sample size, particularly when estimated from the sparse probability bins of survey histograms or the discrete strike prices of option-implied densities. Small perturbations in the tail probabilities can produce large swings in skewness, making it unreliable as a standalone signal \citep{bowley1926elements}.

Second, even when individual skewness estimates are informative, the question of \emph{how much confidence} to assign them remains open. A forecaster reporting moderate positive skewness together with a median close to target sends a weaker directional signal than one reporting the same skewness with a median well above target. The former case may reflect noise; the latter is more likely to reflect a genuine perception of upside risk.

\paragraph{Signal-extraction motivation.}

These obstacles motivate a signal-extraction approach---a formalization of the ``balance of risks''---that combines first- and third-moment information. When $c > 0$, both the median deviation $Q_t - \pi^*$ (a first-moment object, where $\pi^*\equiv\mu^*$) and the asymmetry measure $A_t$ (a third-moment object) should carry the same directional sign. The degree of sign agreement between the two provides a data-driven confidence weight: when both components point in the same direction, the directional signal is strong; when they disagree, the signal is attenuated.

This logic maps directly onto the structure of a coherent detector. In signal processing, a coherent detector exploits the expected alignment between a noisy signal and a reference to distinguish true directional content from noise. Here, $Q_t - \pi^*$ serves as the reference signal (the robust first-moment estimate of the deviation) and $A_t$ serves as the noisy auxiliary signal (the fragile third-moment estimate). Their product $\tilde{Q}_t \tilde{A}_t$ is positive under coherent asymmetry and negative under incoherent conditions---a natural sign-agreement statistic.

\begin{figure}[tbp]
    \centering
    \begin{tabular}{cc}
        \multicolumn{2}{c}{\textbf{COHERENT ASYMMETRY (strong directional signal)}} \\[4pt]
        \textbf{Skewness $>$ 0, Median $>$ Target} & \textbf{Skewness $<$ 0, Median $<$ Target} \\
        \includegraphics[width=0.45\textwidth]{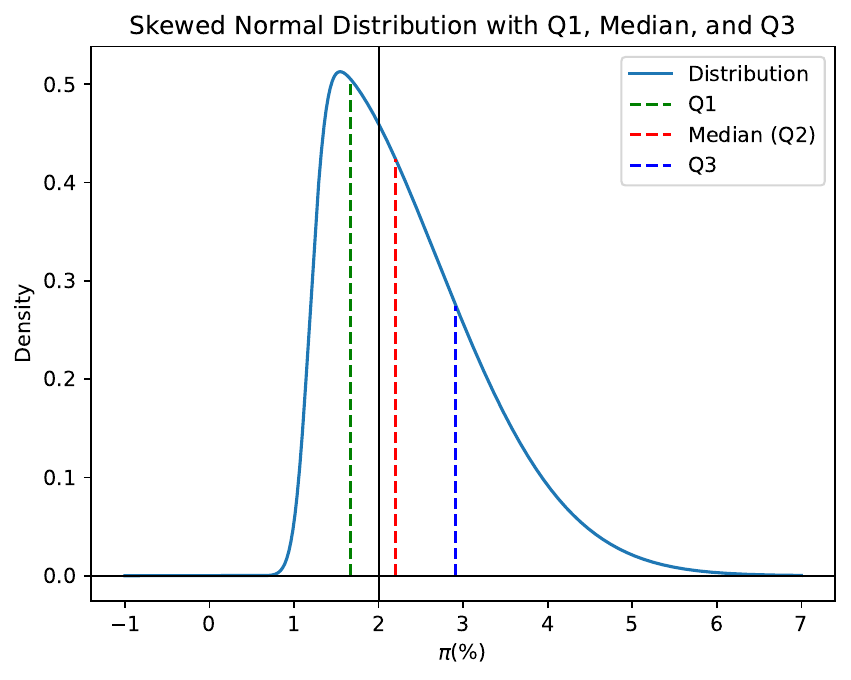} & \includegraphics[width=0.45\textwidth]{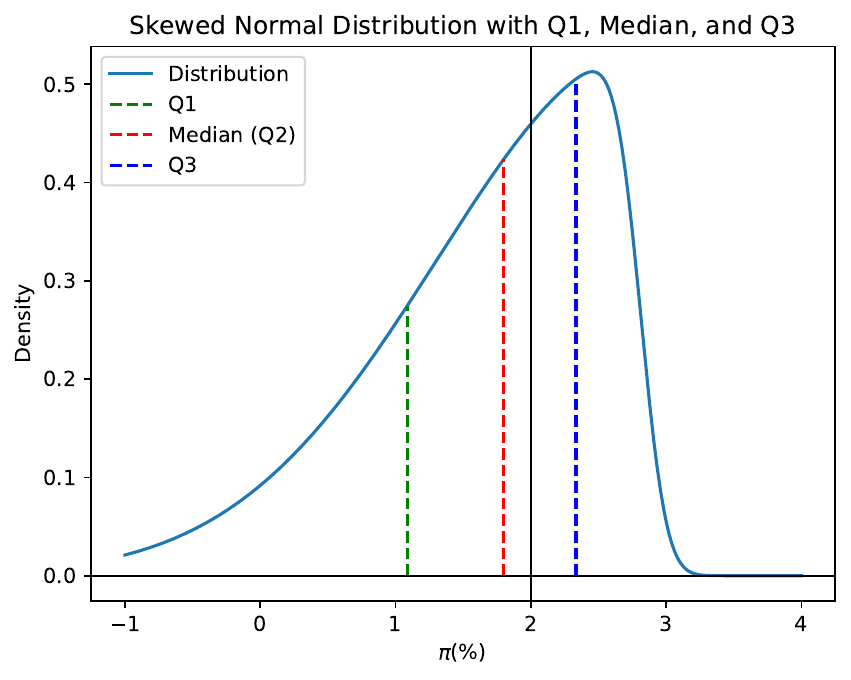} \\[6pt]
        \multicolumn{2}{c}{\textbf{INCOHERENT ASYMMETRY (weak directional signal)}} \\[4pt]
        \textbf{Skewness $>$ 0, Median $<$ Target} & \textbf{Skewness $<$ 0, Median $>$ Target} \\
        \includegraphics[width=0.45\textwidth]{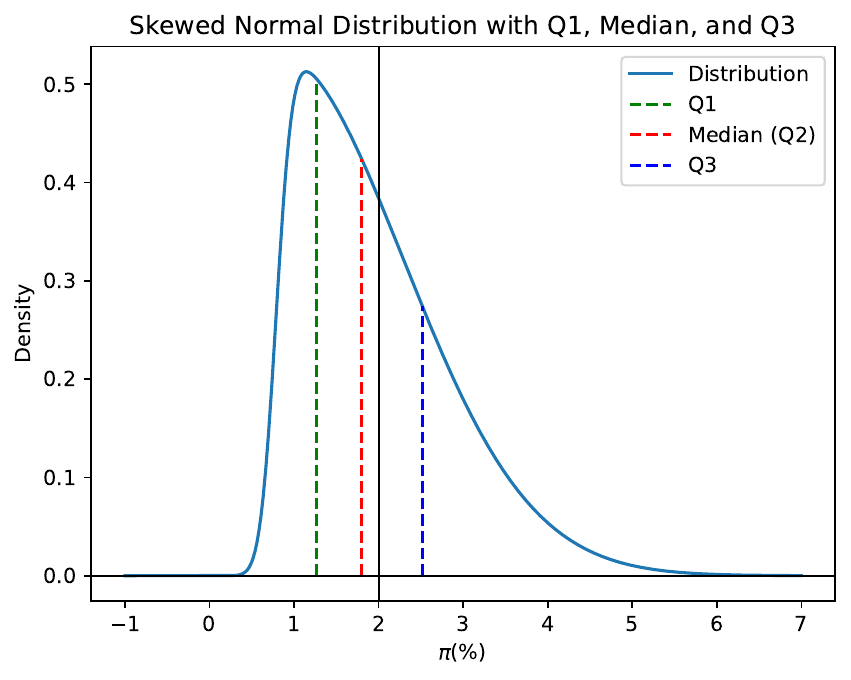} & \includegraphics[width=0.45\textwidth]{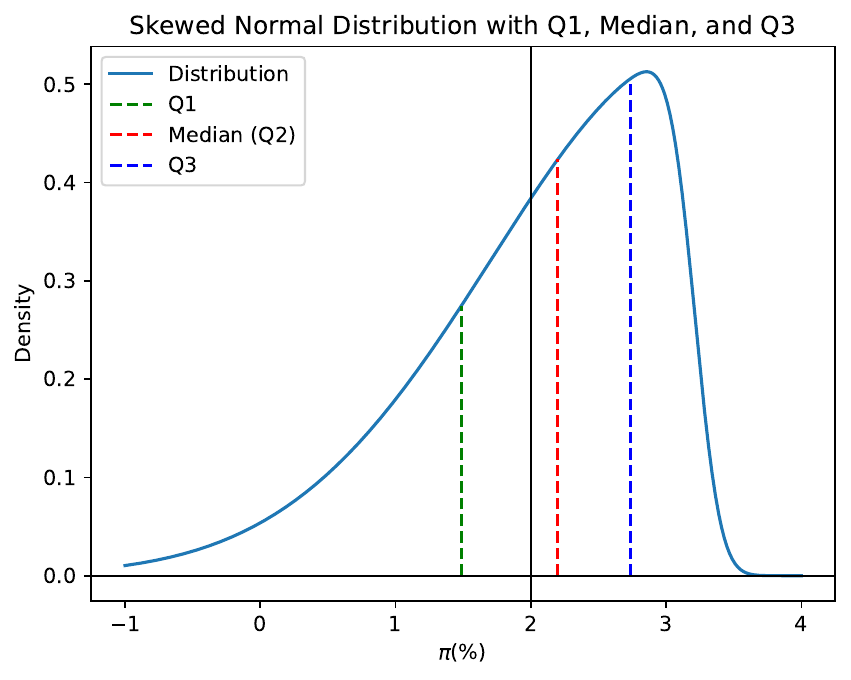} \\
    \end{tabular}
    \caption{Illustration of coherent versus incoherent asymmetry.
    Skew-normal distributions with $Q_1$ (green dashed), median (red dashed),
    and $Q_3$ (blue dashed) marked; the solid black vertical line indicates the
    $2\,\%$ inflation target ($\pi^*$).
    \textbf{Top row:} coherent cases---the sign of skewness agrees with the
    direction of the median deviation from target, producing a strong directional signal.
    \textbf{Bottom row:} incoherent cases---skewness and median deviation point in opposite
    directions, producing a weak or uninformative signal. Script: \texttt{fig07\_skew\_cases.py}.}
    \label{fig:skewed_distributions}
\end{figure}

The strength of a directional signal is assessed through the alignment between the asymmetry of probability distributions and the median's deviation from a reference target. A \textbf{strong signal} emerges when skewness and median deviation indicate the same directional expectation: positive skewness with a median above target (strong upward signal) or negative skewness with a median below target (strong downward signal). Conversely, a \textbf{weak signal} arises from opposing skewness and median deviation: positive skewness paired with a median below target, or negative skewness with a median above target. Figure~\ref{fig:skewed_distributions} illustrates these four scenarios by plotting skew-normal distributions with key quantiles highlighted.

\paragraph{Construction of the AC index.}

The AC index summarizes the directional strength of expectations by jointly exploiting two aggregate statistics extracted from individual SPDs: the cross-sectional median of the forecast distribution, denoted~$Q_t$, and the cross-sectional mean of individual asymmetry scores (computed via Bowley's skewness), denoted~$A_t$.

To ensure comparability across components and robustness to scale differences, both series are normalized using a hyperbolic tangent transformation scaled by their interquartile range (IQR). This normalization preserves signs while bounding magnitudes in~$(-1,1)$:
\begin{equation}\label{eq:ac_normalization}
\tilde Q_t
=
\tanh\!\left(
\frac{Q_t - \pi^\ast}{\mathrm{IQR}(Q_t - \pi^\ast)}
\right),
\qquad
\tilde A_t
=
\tanh\!\left(
\frac{A_t}{\mathrm{IQR}(A_t)}
\right),
\end{equation}
where $\pi^\ast$ denotes the reference inflation target (equal to $\mu^*$ in the notation of Section~\ref{sec:distance_asymmetry}).

The AC index combines these two normalized components through a smooth coherence-based aggregator:
\begin{equation}\label{eq:ac}
\mathrm{AC}_t
=
\underbrace{
\left(
\frac{\tilde Q_t + \tilde A_t}{2}
\right)}_{\text{directional signal}}
\underbrace{
\left(
\frac{1 + \tilde Q_t \tilde A_t}{2}
\right)}_{\text{coherence weight}},
\end{equation}
which is bounded in~$[-1,1]$.

\begin{figure}[tbp]
  \centering
  \includegraphics[width=\linewidth]{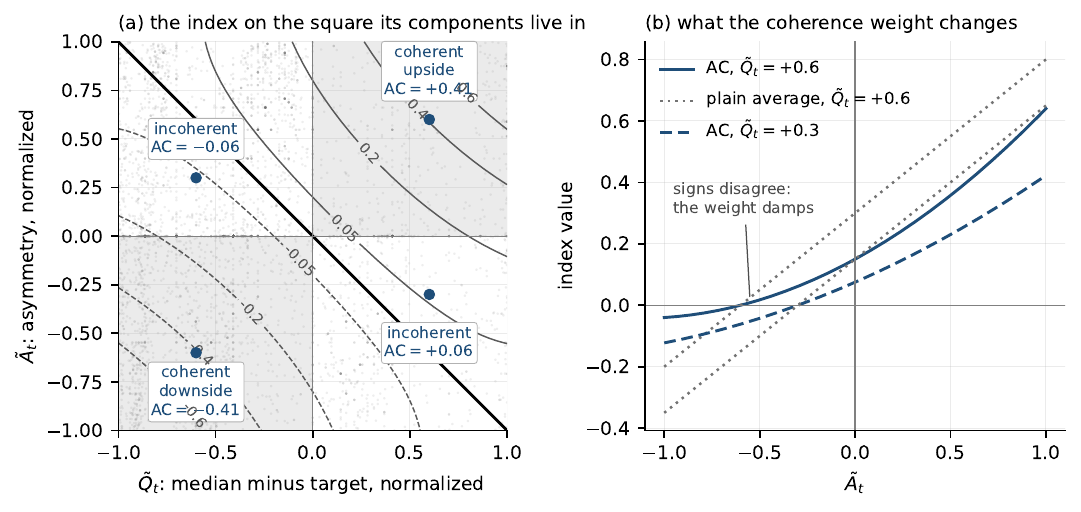}
  \caption{\textbf{What the coherence weight does.} \emph{Panel (a):} contours of
  $\mathrm{AC}$ of~\eqref{eq:ac} over the square its two normalized components live in, the zero contour
  in bold and the quadrants where the components agree in sign shaded; the dots are the individual
  densities of the panel, and the four marked points are the cases of
  Figure~\ref{fig:skewed_distributions}. \emph{Panel (b):} the index against the plain average of the two
  components, as the asymmetry runs across its range at two distances from target.}
  \label{fig:ac_map}
  \small
  \textit{Note:} ECB-SPF one-year-ahead inflation densities, 1999Q1--2026Q3, $4{,}638$ forecaster-rounds;
  $\tilde Q_t$ and $\tilde A_t$ are built as in~\eqref{eq:ac_normalization}, the asymmetry smoothed over
  two rounds by forecaster. $47\,\%$ of the densities report an asymmetry whose sign differs from that of
  their own median gap, which is what the shaded and unshaded quadrants separate. Script:
  \texttt{fig08\_ac\_map.py}.
\end{figure}

Figure~\ref{fig:ac_map} draws it over the square the two components live in. The contours turn with the
coherence weight: at the same distance from target, a round whose asymmetry agrees in sign with that
distance scores about seven times what a round of equal magnitudes and opposite signs scores---$\pm 0.41$
against $\pm 0.06$ at $|\tilde Q_t|=0.6$---and the plain average of the two components, drawn beside the
index, makes no such distinction. The cloud of individual densities behind the contours shows that this is
not a corner case: $47\,\%$ of them report an asymmetry whose sign differs from that of their own median
gap.

\paragraph{Structural interpretation.}

The two factors in~\eqref{eq:ac} have transparent structural interpretations in the light of the reduced form~\eqref{eq:skew_distance_preview}.

The first factor, $(\tilde Q_t + \tilde A_t)/2$, is the \emph{signed directional signal}. Under coherent asymmetry ($c > 0$), both $\tilde Q_t$ and $\tilde A_t$ should share the sign of $\mu_t - \mu^*$, so their average estimates the direction of the deviation. Taking the average rather than either component alone reduces the variance of the estimator by pooling two noisy proxies for the same latent direction.

The second factor, $(1 + \tilde Q_t \tilde A_t)/2$, is the \emph{coherence weight}. It isolates the extent to which the first and third moments convey a consistent directional message. When both components share the same sign, $\tilde Q_t \tilde A_t > 0$ and the weight exceeds~$\frac{1}{2}$, amplifying the directional signal. When they have opposite signs---indicating that the skewness contradicts the median deviation---the weight falls below~$\frac{1}{2}$, attenuating the signal. Under perfect anti-coherence ($\tilde Q_t \tilde A_t = -1$), the weight collapses to zero and $\mathrm{AC}_t = 0$ regardless of the magnitudes.

For interpretability, the coherence component is also reported separately:
\begin{equation}\label{eq:coherence}
\mathrm{Coherence}_t
=
\frac{1 + \tilde Q_t \tilde A_t}{2},
\end{equation}
which lies in~$[0,1]$ and provides a direct diagnostic of whether the two moments are aligned.

\paragraph{Formal properties.}

The AC index satisfies several desirable properties:
\begin{enumerate}[label=(\roman*)]
  \item \emph{Boundedness}: $\mathrm{AC}_t \in [-1,1]$ for all $(\tilde Q_t, \tilde A_t) \in [-1,1]^2$.
  \item \emph{Symmetry}: $\mathrm{AC}_t(-\tilde Q, -\tilde A) = -\mathrm{AC}_t(\tilde Q, \tilde A)$; reversing the signs of both components reverses the direction of the index.
  \item \emph{Coherence monotonicity}: for fixed $|\tilde Q_t|$ and $|\tilde A_t|$, $|\mathrm{AC}_t|$ is increasing in $\tilde Q_t \tilde A_t$; greater sign agreement amplifies the signal.
\end{enumerate}
Section~\ref{sec:simulation} provides Monte Carlo evidence that the AC formula outperforms alternative aggregation rules---including simple averages and skewness-only indicators---in recovering the true directional state under realistic noise calibrations.

\begin{remark}[Portability and the balance of risks]
\label{rem:ac_portability}
The AC construction is not specific to survey histograms.
It applies to any predictive distribution for which one can define (i)~a reference point for the central tendency and (ii)~a measure of distributional asymmetry.
In option-implied densities, the forward price provides the reference and implied skewness the asymmetry signal; AC would then measure whether downside or upside tail risk is coherently aligned with the expected path of the underlying, filtering out smile-interpolation artifacts and maturity-dependent distortions.
In firm-level expectation surveys, industry benchmarks or historical means serve as anchors.
In GDP growth forecasts, potential output provides the reference.
More broadly, AC operationalizes the concept of \emph{balance of risks}---a notion ubiquitous in central-bank communication, financial stability assessments, and risk management, yet lacking a standard statistical formalization.
AC is not merely ``better skewness'': it separates shape asymmetry (a statistical property) from decision-relevant directional asymmetry (an economic property), a distinction that matters in any setting where practitioners care about which tail of a distribution is more consequential.
\end{remark}

\begin{remark}[Dependence on first-moment quality]
\label{rem:ac_firstmoment}
The signal-extraction power of AC rests on the premise that the first-moment deviation $Q_t - \pi^*$ is observed more reliably than the third-moment signal $A_t$.
This premise is well founded in standard inflation-targeting regimes, where the target $\pi^*$ is publicly announced and the cross-sectional median $Q_t$ is robustly estimated from a large panel.
In low-credibility environments where the anchor itself is ambiguous---for instance, when the central bank lacks an explicit target or when the target is widely believed to be non-binding---the coherence weight $|\tilde{Q}_t|$ becomes a noisy proxy, and the denoising advantage of AC over raw asymmetry may degrade.
In the ECB sample, the two-percent target provides a stable, publicly known reference throughout the estimation period; the AC construction thus operates in a favorable identification environment.
\end{remark}

\subsection{The series, and its distinctness from Normalized Uncertainty}
\label{sec:AC_empirical}

\paragraph{Validating the skewness--deviation relationship.}

The reduced form~\eqref{eq:skew_distance_preview} predicts a linear relationship between skewness and the signed deviation from the anchor. The regression estimated is
\begin{equation}
  \label{eq:skew_regression}
  A_{i,t} = \alpha_0 + \alpha_1\,(\mu_{i,t} - \mu^*) + u_{i,t}
\end{equation}
on individual ECB-SPF inflation SPDs, where $A_{i,t}$ is Bowley's skewness for forecaster~$i$ in survey round~$t$ and $\mu_{i,t}$ is the corresponding SPD mean. A positive and significant $\hat\alpha_1$ confirms that asymmetry is coherent on average, $c>0$.

Table~\ref{tab:skew_dev_regs} reports four specifications. In the baseline the slope is
$\hat\alpha_1=0.0128$ ($t=7.0$, standard errors clustered by survey round): a forecaster who expects
inflation further above target reports a more positively skewed density, and symmetrically below it, which
is the positive mean--skewness relation of Table~\ref{tab:momentscorr} at the level of the individual
density. The slope survives what is asked of it. It rises to $0.0147$ ($t=4.2$) once survey-round effects
absorb everything common to a round, so it is identified from disagreement within a round and not from
shared time variation; it is $0.0126$ ($t=4.4$) when the deviations beyond two points are trimmed, and
$0.0103$ ($t=2.3$) when the trimming and the round effects are imposed together. What the relation does
not do is explain much: the $R^2$ is $0.011$ in the baseline and $0.056$ with round effects. A sign that
holds across every specification and a share of explained variation of a few per cent are exactly the two
facts the construction rests on --- it needs the sign of $c$ in~\eqref{eq:skew_distance_preview}, which is
positive, and it is built for the noise that the small $R^2$ measures.

\paragraph{The euro-area series.}

The growth counterpart, ACG, is built in Section~\ref{sec:growth}.

\begin{figure}[tbp]
  \centering
  \includegraphics[width=\linewidth]{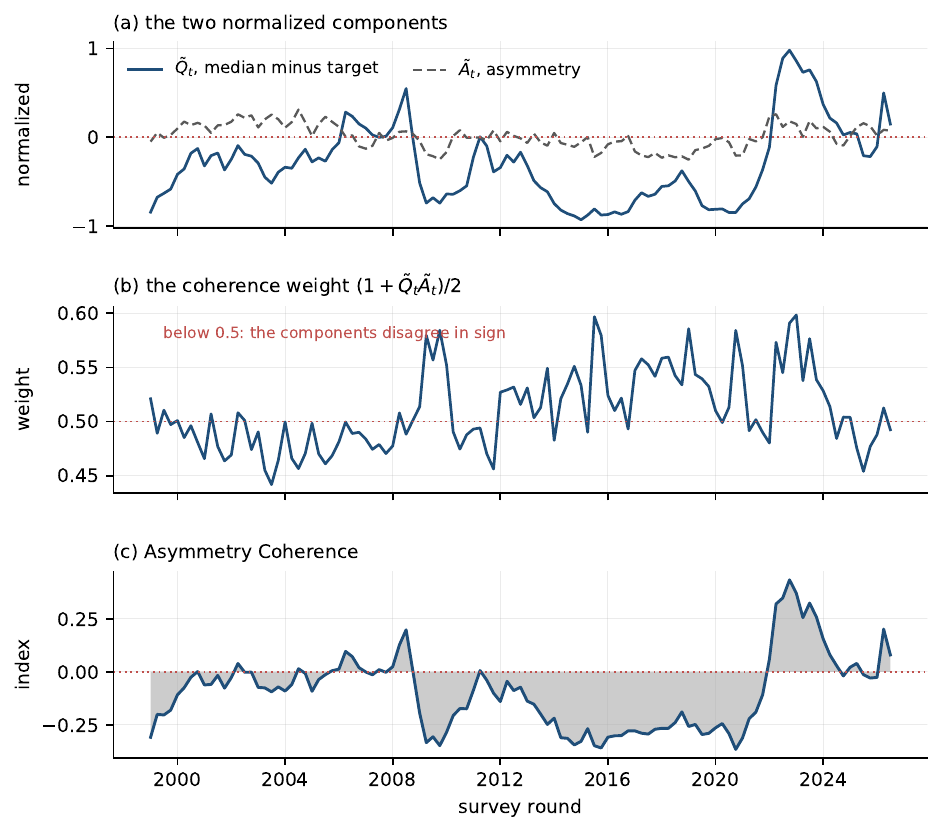}
  \caption{\textbf{Asymmetry Coherence and the pieces it is made of, ECB-SPF one-year-ahead inflation.}
  \emph{Panel (a):} the two normalized components, the median's distance from target $\tilde Q_t$ and the
  asymmetry $\tilde A_t$. \emph{Panel (b):} the coherence weight $(1+\tilde Q_t\tilde A_t)/2$, which is
  below one half exactly when the two disagree in sign. \emph{Panel (c):} the index itself. All three are
  cross-forecaster means of the individual values, on the round's own quarter.}
  \label{fig:ac_series}
  \small
  \textit{Note:} $4{,}638$ forecaster-rounds over $111$ rounds. The index is negative in $82$ rounds and
  positive in $29$, and exceeds $0.1$ in absolute value in $62$. The weight reads mild at the round level,
  between $0.44$ and $0.60$, because disagreement across forecasters averages out; on the individual
  densities where it is applied it runs from $0.19$ to $0.82$ between the fifth and the ninety-fifth
  percentile. Script: \texttt{fig09\_ac\_series.py}.
\end{figure}

Figure~\ref{fig:ac_series} shows the AC index for one-year-ahead inflation forecasts from the ECB-SPF. Prior to the Global Financial Crisis, the first-moment component~$\tilde Q_t$ and the asymmetry component~$\tilde A_t$ display little systematic coherence, indicating weak directional signals in inflation expectations. During the Zero Lower Bound (ZLB) period, forecasters exhibit a pronounced and persistent downside signal, reflecting sustained concerns about below-target inflation. By contrast, during the recent euro-area inflation surge, expectations initially display strong upside risk signals, which gradually dissipate by the end of 2023. Notably, changes in distributional asymmetry anticipate shifts in the median forecast, suggesting that skewness contains early information about emerging inflation risks. The three panels separate the pieces: the two normalized components, the coherence weight that multiplies their average, and the index. The weight reads mild at the round level, moving between $0.44$ and $0.60$, because disagreement across forecasters averages out; at the level where it is applied, the individual density, it spans almost the whole unit interval, from $0.19$ to $0.82$ between the fifth and the ninety-fifth percentile.

\paragraph{Distinctness from Normalized Uncertainty.}

Over the euro-area sample the AC index is only moderately related to the asymmetry component
it is built from --- the correlation with the round-mean Bowley skewness is $0.53$, so
roughly a quarter of its variation is skewness --- and closely related to the median
component, $0.92$. That is the construction working as designed: the coherence weight lets
the first moment carry the direction and the third moment qualify it.

Against Normalized Uncertainty the answer depends on which NU series is used, and the
dependence is itself informative. Regressing the standardized AC index on standardized NU
gives $+0.16$ ($t=1.6$, $R^2=0.02$) with the calibrated series and $-0.13$ ($t=-1.4$,
$R^2=0.02$) with the fitted one. The ordering is not an instability but a mechanism, and it
follows what the denominator removes. The directional signal is high when the forecast sits
above target and low when it sits below; the normalization deflates above target only, which
is exactly where that signal is, so the more the denominator deflates there the less the
corrected series shares with AC --- until, with the ratio matched to the envelope, nothing
measurable is left. Under every calibration the shared variation stays below a third. AC is
therefore not a repackaging of NU, whichever NU is chosen; what the sign of their overlap says
about an economy is a question this paper poses and does not answer.

    \section{The growth measures, and the orthogonalization}
\label{sec:growth}

Nothing in the construction of Section~\ref{sec:nu} is specific to inflation. It needs a
reference point and a dispersion that moves with the distance from it. The ECB survey
collects density forecasts for real GDP growth on the same rounds and from the same
forecasters, which makes growth the natural place to show what the construction requires
and what has to change when one ingredient --- an announced number --- is missing. The
measures built here are used in the companion paper to ask whether what is priced is
uncertainty about inflation or uncertainty in general; that question is what
Section~\ref{sec:orth} equips.

\subsection{The reference point}
\label{sec:gpot}

Growth has a benchmark but no announced target. The reference used here is potential
growth $g^{\mathrm{pot}}_t$, measured as the annualized growth rate of the
Hodrick--Prescott trend of log real euro-area GDP,
\begin{equation}
  \label{eq:gpot}
  g^{\mathrm{pot}}_t \;=\; 400\,\bigl(\log \widehat{Y}_t - \log \widehat{Y}_{t-1}\bigr),
\end{equation}
with $\widehat{Y}$ the trend component at the quarterly smoothing constant
$\lambda=1600$, computed on chain-linked volumes, aligned to the round in which the
forecast was formed and carried forward where the national-accounts release lags the
survey. The trend is computed on the full available sample, so the value of
$g^{\mathrm{pot}}_t$ at a given date is not the value a forecaster could have computed at
that date; this is a filter, not a real-time benchmark, and the end-point sensitivity of
the Hodrick--Prescott trend falls on the most recent rounds.

The substantive difference from Section~\ref{sec:nu} is not the filter. It is that
$g^{\mathrm{pot}}_t$ is an estimate and $\mu^{*}=2\,\%$ is an announcement. A kink can sit
at an announced number because everyone can see it; there is no comparable point on the
growth axis, so the growth denominator is symmetric by construction, as the next subsection
states.

\subsection{Normalized Growth Uncertainty}
\label{sec:ngu}

From the one-year-ahead real GDP growth densities, with the mean $\mu_{i,t}$ and the
standard deviation $\sigma_{i,t}$ of forecaster $i$'s reported histogram computed under the
conventions of Section~\ref{sec:moments},
\begin{equation}
  \label{eq:ngu}
  \mathrm{NGU}_{i,t}
  \;=\; \frac{\sigma_{i,t}}{\sqrt{1+\bigl|\mu_{i,t}-g^{\mathrm{pot}}_t\bigr|}},
  \qquad
  \mathrm{NGU}_t \;=\; \frac{1}{N_t}\sum_{i=1}^{N_t}\mathrm{NGU}_{i,t},
\end{equation}
the round series being the mean of the individual values, as everywhere in this paper. Two
choices are deliberate. The denominator is \emph{symmetric}: with no announced number there
is no point at which a kink could sit, so the one-sided form of~\eqref{eq:nu_cal} is not
imposed. And the coefficients are set to unity rather than fitted, so that the series is
reproducible from the histograms and the trend alone --- the calibrated choice
of~\eqref{eq:nu_cal}, for the same reason. The series runs from 1999Q1 to the latest round.

\begin{figure}[tbp]
  \centering
  \includegraphics[width=0.95\linewidth]{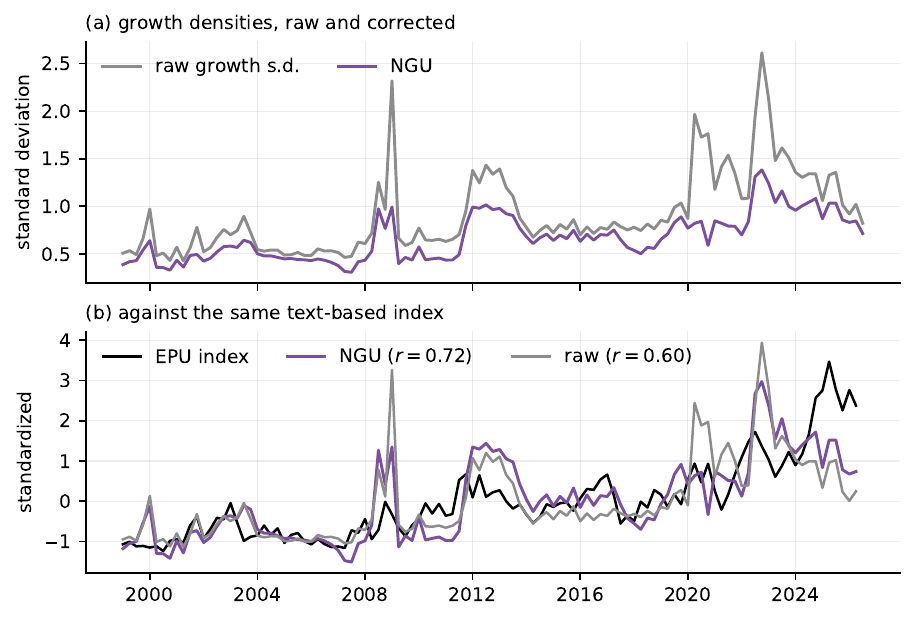}
  \caption{\textbf{Normalized Growth Uncertainty.} The raw round-mean standard deviation of the
  one-year-ahead growth densities and its corrected counterpart~\eqref{eq:ngu}, and the
  corrected series against the euro-area economic policy uncertainty index. Script: \texttt{fig10\_ngu\_series.py}.}
  \label{fig:ngu_series}
\end{figure}

Figure~\ref{fig:ngu_series} draws it against its raw counterpart and against the euro-area
economic policy uncertainty index. The correction raises the agreement with that index from
$0.601$ to $0.725$ over the 110 quarters the two overlap. Agreement with an independent proxy \emph{rising} under a
correction is the pattern a measurement fix should produce and a difficult one to obtain by
accident: the text-based index shares none of the survey's construction, so nothing
mechanical links the two.

The inflation and growth series are read together outside the academic literature as well:
Banque de France policy communication uses the pair to report that professional forecasters
continue to express high levels of uncertainty about both variables
\citep{vansteenberghe2026ecoblog}. That is the use the calibrated construction is meant to
support --- a series someone else can rebuild from the published histograms and the trend,
without first having to estimate an envelope.

\subsection{Asymmetry Coherence of Growth}
\label{sec:acg}

The asymmetry construction of Section~\ref{sec:AC_measure} carries over unchanged except
for the reference point. With $Q^{g}_t$ the cross-forecaster median of the individual
growth-density medians and $A^{g}_t$ the round mean of the individual Bowley skewness
values,
\begin{equation}
  \label{eq:acg}
  \tilde Q^{g}_t=\tanh\!\left(\frac{Q^{g}_t-g^{\mathrm{pot}}_t}
                                   {\mathrm{IQR}(Q^{g}_t-g^{\mathrm{pot}}_t)}\right),
  \qquad
  \tilde A^{g}_t=\tanh\!\left(\frac{A^{g}_t}{\mathrm{IQR}(A^{g}_t)}\right),
\end{equation}
and $\mathrm{ACG}_t$ is~\eqref{eq:ac} evaluated at $(\tilde Q^{g}_t,\tilde A^{g}_t)$. The
caution of Remark~\ref{rem:ac_firstmoment} binds harder here than for inflation: the
coherence weight is only as good as the reference, and a filtered trend is a noisier
reference than an announced target.

\subsection{Inflation-specific or general uncertainty: the orthogonalization}
\label{sec:orth}

A user who wants the part of inflation uncertainty that is not general uncertainty --- or
the reverse --- cannot get it by regressing one round series on the other. The two measures
are built forecaster by forecaster, and forecasters differ in how widely they report:
a round-level regression would remove, along with the common component, the composition of
the panel that quarter. The purge is therefore done where the measures are made.

On the panel of forecaster-rounds carrying both an inflation and a growth density
($4{,}327$ forecaster-rounds through 2026Q2, $4{,}350$ through the latest round), for each
forecaster $i$ with at least ten matched rounds,
\begin{equation}
  \label{eq:orth}
  \mathrm{NGU}_{i,t} \;=\; a_i + b_i\,\mathrm{NU}_{i,t} + e_{i,t},
  \qquad
  (\mathrm{NGU}\!\perp\!\mathrm{NU})_{i,t} \;\equiv\; e_{i,t},
\end{equation}
is estimated on that forecaster's own rounds --- own intercept and own slope, the strictest
reading of ``forecaster by forecaster'' ($86$ forecasters). The remaining $21$, who report
too few matched rounds to carry a slope of their own, take the within-forecaster slope with
their own intercept. The residuals are averaged within each round and the round series is
standardized. The symmetric construction, regressing $\mathrm{NU}$ on $\mathrm{NGU}$, gives
$(\mathrm{NU}\!\perp\!\mathrm{NGU})$.

\begin{figure}[tbp]
  \centering
  \includegraphics[width=0.95\linewidth]{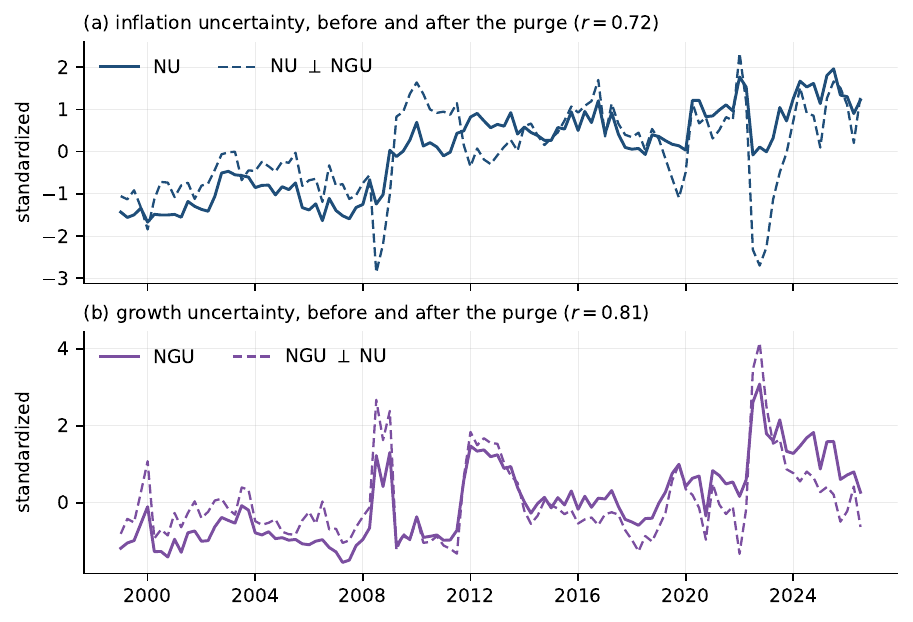}
  \caption{\textbf{Inflation and growth uncertainty, before and after the purge.}
  \emph{Panel (a):} the round series of Normalized Uncertainty and the same series purged of growth
  uncertainty. \emph{Panel (b):} Normalized Growth Uncertainty and the same series purged of inflation
  uncertainty. The purge is the forecaster-by-forecaster orthogonalization of~\eqref{eq:orth}; every
  series is standardized over its own sample, and each panel's title gives the correlation between the
  two lines it draws. Script: \texttt{fig11\_nu\_ngu\_orthogonalization.py}.}
  \label{fig:nu_ngu_orth}
\end{figure}

Which inflation series enters matters and is stated: the companion paper's regressions use
$\mathrm{NU}^{\mathrm{fit}}$, so that is the baseline here, with
$\mathrm{NU}^{\mathrm{cal}}$ reported as a check; the two orthogonalized series correlate at $0.99$. Table~\ref{tab:ngu_orth} reports the panel counts, the correlation between the
two measures and the agreement of each series with the text-based index. Over the full
sample, 1999Q1--2026Q3, inflation and growth uncertainty move together, $+0.66$.
Figure~\ref{fig:nu_ngu_orth} draws each series beside its purged counterpart: the purge moves the inflation measure more than the growth one, the correlation between a series and its residual being $0.72$ against $0.81$.

The orthogonalized components are measures, not identification. Nothing in the construction
says that the surviving component is causal for anything; it says only that the part of one
measure that the other predicts, forecaster by forecaster, has been taken out. What that
buys is a test that an applied paper can run: \citet{vansteenberghe2026tolerable} uses these
series to ask whether bank loan pricing responds to uncertainty about inflation or to
uncertainty in general.
    \section{Carrying the constructions to another survey}
\label{sec:other_surveys}

The measures above were built on one survey. Two things have to be settled before they can
be built on a second: what the other survey's grid does to a measured variance, and what may
be compared across the two once each has been measured on its own terms. The US Survey of
Professional Forecasters, run by the Federal Reserve Bank of Philadelphia, is the useful
case, because it has something the euro-area survey does not --- an announcement of a
numerical inflation objective inside the sample, in January 2012 --- and because it has a
change of histogram grid two years later that a careless reading attributes to the
announcement.

\subsection{The US density questions and their grids}
\label{sec:us_grids}

The survey asks for densities over three price variables, and they do not share a history.
The oldest question, on the GDP price index, runs from 1992 and has the long record on both
sides of the announcement --- but its grid halves its bin widths at 2014Q1, from ten
one-point bins to ten half-point bins. The core PCE and core CPI questions begin in 2007Q1
and their grid has never changed. In every case the workbook reports two blocks of ten
probabilities, the current year and the next, and the one-year-ahead object is the second
block; the columns run from the highest bin down, which is the reverse of the euro-area
layout. One-decimal reporting makes a
bin labelled ``3.5 to 3.9'' the interval $[3.5,4.0)$, so its midpoint is $3.75$, and the two
open tails are given one bin width. Moments are computed as in
Section~\ref{sec:moments}; because the two grids differ in width, each variance is also
reported with Sheppard's correction. The reference point is
$\mu^{*}=2\,\%$ throughout, which before January 2012 is a retrospective convention and is
used as such.

\subsection{A break that is a change of grid}
\label{sec:us_artefact}

\paragraph{Why the GDP-price question cannot answer it.} The survey's oldest density
question, on the GDP price index, spans the announcement with a long history on either
side, and a naive reading of it appears to deliver a large result: the variance--distance
slope falls from $0.291$ to $0.072$ across 2012Q1, with a Chow statistic above $100$.
That break is a measurement artefact. The histogram grid for this question halves its bin
widths at 2014Q1, mean measured variance falls from $0.740$ to $0.346$ across that date,
and $1{,}446$ of the $1{,}744$ post-2012 observations sit in the new grid --- so the
"post-announcement" sample is very nearly the "post-grid-change" sample.

Running the two events against each other settles it. The grid interaction is $-0.413$
while the announcement interaction is $+0.183$ with a $t$ of $0.73$: wrong-signed and
insignificant. Within the constant grid the slope moves from $0.291$ to $0.474$,
wrong-signed again. In arms form the grid interaction on the overshoot arm is $-0.679$
($t=-2.87$) against $+0.427$ for the announcement. Eight quarters separate the two events
and this design cannot separate them. The figures $0.291\to0.072$ are therefore not an
announcement effect and are used in neither paper.

\subsection{The question on a grid-stable series, and what may be compared}
\label{sec:us_rule}

\paragraph{The grid-stable questions.} The obstacle is specific to
the GDP-price question. The survey has asked for densities over core PCE and core CPI
inflation since 2007Q1 on a grid that has \emph{never} changed, which places the January
2012 announcement inside a window with twenty quarters before it, fifty-eight after, and
one measurement convention throughout. \citet{vansteenberghe2026tolerable} reports the result: the
above-target arm steepens by $+0.924$ ($t=4.20$) on core CPI and $+0.524$ ($t=2.12$) on
core PCE, against below-target arms of $+0.448$ ($t=2.9$) and $-0.032$ ($t=-0.2$) --- so
the below-target side follows at half the magnitude on core CPI and not at all on core
PCE --- and the hypothesis that the announcement moved both arms equally is rejected at
$p=0.001$ and $p=0.013$. Those $p$-values come from a Wald test on the joint covariance of
the two arm interactions, not from a naive difference of independent estimates.

\paragraph{The unit rule.} What remains is the cross-source comparison itself, and here no slope is compared across the two surveys: their price indices differ, their horizons differ, and their density grids differ. What is comparable is the \emph{shape}---flat below, rising above---and the ratio $b_+/a$ within each source. On that basis the euro area's individual densities return $2.61$ and the US individual-level core-CPI series $0.20$, an order of magnitude apart, and \citet{vansteenberghe2026tolerable} sets out the three candidate reasons without testing any of them.

The rule this leaves is general, and it is the one result of this section that travels to
any survey: compare shapes and within-source ratios, never slopes across sources, and treat
a change of grid as a standing control on every estimate the changed question enters.
Table~\ref{tab:us_arms} reports the arms by question, with and without that control.

\subsection{Asymmetry across surveys}
\label{sec:us_asymmetry}

The coherence regression of Section~\ref{sec:AC_empirical} can be run on the US densities
in the same way, on the grid-stable core questions and with the grid change controlled for
--- and since the core grid never changed, no control is called for. The answer is that the
euro-area sign does not travel. Before the announcement the slope is negative on both core
questions, $-0.021$ ($t=-2.3$) on core CPI and $-0.024$ ($t=-3.7$) on core PCE: US densities
were \emph{anti}-coherent, forecasters expecting inflation above target placing relatively
more mass below it, which is what an anticipated policy correction looks like. After the
announcement the slope is indistinguishable from zero on both questions, $-0.007$ ($t=-0.9$)
and $+0.005$ ($t=0.8$), against $+0.013$ ($t=7.0$) in the euro area over its whole sample.
Section~\ref{sec:distance_asymmetry} said that the sign of~$c$ is an empirical question,
with coherent and anti-coherent channels pulling against each other; two surveys answering
it differently is the strongest evidence here that the sign belongs to an institutional
setting rather than to survey densities as such. The earlier cross-country comparison, run
on the GDP-price question, is not used, for the reason
Section~\ref{sec:us_artefact} gives.

\subsection{The two blocks side by side}
\label{sec:us_side_by_side}

\begin{figure}[tbp]
  \centering
  \includegraphics[width=\linewidth]{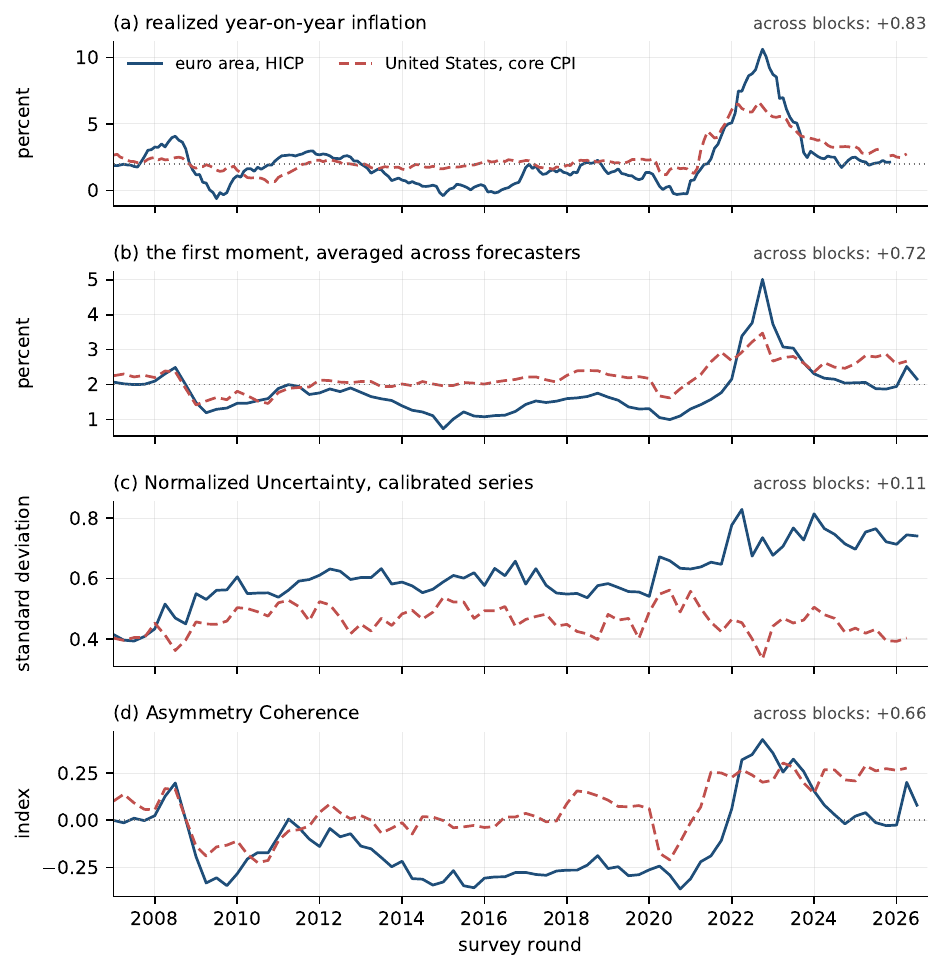}
  \caption{\textbf{The euro area and the United States, round by round.} \emph{Panel (a):} realized
  year-on-year inflation, each block's own target variable. \emph{Panel (b):} the first moment, the
  cross-forecaster mean of the individual density means. \emph{Panel (c):} Normalized Uncertainty in
  its calibrated form. \emph{Panel (d):} Asymmetry Coherence, each block normalized on its own sample.
  The euro area is drawn solid and the United States dashed; the dotted horizontal lines are the target
  in~(a) and~(b) and symmetry in~(d). Each panel's title gives the correlation between the two series
  it draws.}
  \label{fig:ea_us}
  \small
  \textit{Note:} 2007Q1--2026Q2, the $78$ rounds the two surveys share, with $30$ to $52$ euro-area and
  $22$ to $43$ US forecasters per round. The euro-area question is one-year-ahead headline HICP and the
  grid-stable US question one-year-ahead core CPI, the US survey carrying no headline-CPI density; the
  levels of the two blocks are therefore not comparable, and the figure is read for co-movement only.
  Both blocks are measured with the constructions of Sections~\ref{sec:moments}, \ref{sec:nu}
  and~\ref{sec:AC_measure}. Script: \texttt{fig13\_ea\_us\_comparison.py}.
\end{figure}

The unit rule of Section~\ref{sec:us_rule} forbids comparing slopes across surveys; it does not forbid
asking whether the measured series move together. Figure~\ref{fig:ea_us} places the four objects of this
paper side by side over the $78$ rounds the two surveys share, and the answer differs sharply from one
object to the next. Realized inflation correlates at $+0.83$ across the two blocks and the first moment
at $+0.72$: the level of expected inflation is close to a common object, through the disinflation of the
2010s and through the surge of 2021--23 alike. Asymmetry Coherence follows at $+0.66$. Normalized
Uncertainty does not follow at all. Its cross-block correlation is $+0.11$, and panel~(c) shows the
shape of the divergence: the euro-area series steps up in 2021 and stays up, while the US series returns
to the range it occupied before.

That ordering does not depend on which US price question is used or on where the reference point is put.
Rebuilding the US block on core PCE returns $+0.77$ for the first moment, $+0.13$ for Normalized
Uncertainty and $+0.75$ for Asymmetry Coherence; moving the reference point to $2.3\,\%$, which allows
for the wedge between the CPI and the PCE index on which the US objective is defined, returns $+0.72$,
$+0.23$ and $+0.73$. In all three constructions the normalized second moment travels across the two
blocks far less than the first moment or the asymmetry does.

Two readings the figure does not support should be set aside. The gap in the level of Normalized
Uncertainty, $0.61$ in the euro area against $0.46$ in the United States, is not a difference in
perceived uncertainty: the two questions are asked about different indices on different grids, which is
the comparison the unit rule excludes. And the level of the US coherence index changes sign with the
reference point, $+0.06$ at $2\,\%$ and $-0.06$ at $2.3\,\%$, so nothing here establishes that
euro-area densities are more downside-tilted than US ones. What survives is a statement about
co-movement, and it is the one this section needs: the first moment is close to a common object across
the two blocks, and normalized uncertainty is close to a domestic one.

\subsection{What the construction needs elsewhere}
\label{sec:portability}

Beyond surveys, the requirements are short. Normalized Uncertainty needs a reference point
that the dispersion of the predictive distribution is measured against, and a dispersion
that moves with the distance from it; where the reference is announced, the envelope may be
one-sided, and where it is estimated, as for growth in Section~\ref{sec:growth}, it should
not be. Asymmetry Coherence needs a central tendency measured more reliably than the
asymmetry, and a bounded asymmetry statistic. Option-implied densities meet both, with the
forward price as the reference; firm-level expectation surveys meet both, with an industry
benchmark or a historical mean. What no setting excuses is the measurement discipline of
Section~\ref{sec:data}: the grid, the tails and the level at which the statistics are formed
decide what the moments mean before any normalization is applied to them.
    \section{Validation by simulation}
\label{sec:simulation}

\begin{figure}[tbp]
  \centering
  \includegraphics[width=\linewidth]{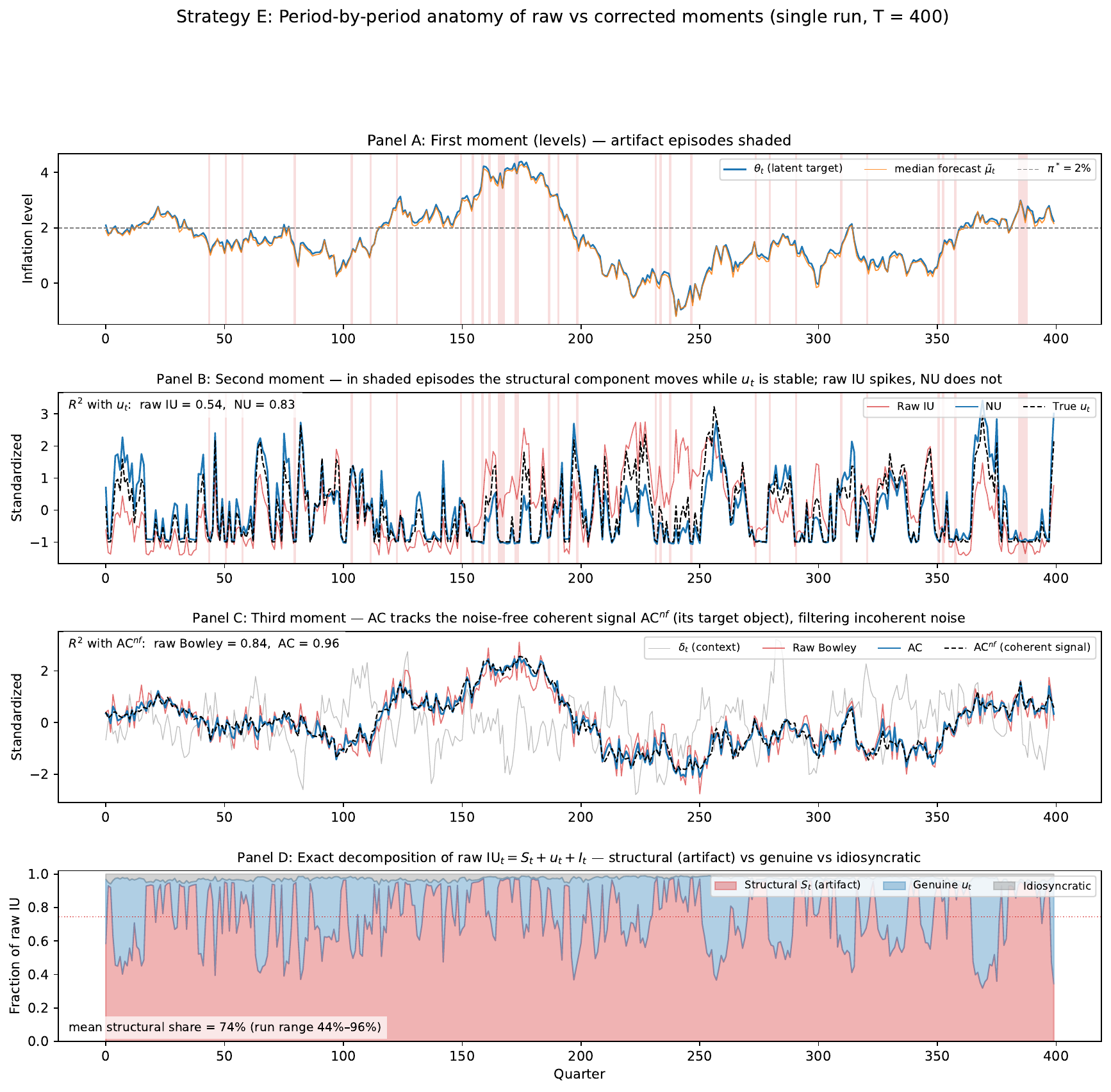}
  \caption{Period-by-period decomposition of raw and corrected moments
  (Monte Carlo, single extended run, $T=400$).
  \textbf{Panel~A:} Latent inflation target $\theta_t$ relative to the anchor $\pi^*$
  and the median forecast $\tilde\mu_t$, in levels; the shaded episodes are those in which the
  structural component moves sharply while genuine uncertainty $u_t$ is stable.
  \textbf{Panel~B:} Raw IU captures the artifact---it rises whenever $\theta_t$ drifts
  from $\pi^*$ even when genuine uncertainty $u_t$ is stable---while NU tracks the true $u_t$ closely.
  \textbf{Panel~C:} AC filters incoherent noise from raw asymmetry, tracking the noise-free
  coherent signal $\mathrm{AC}^{nf}$; the true directional risk $\delta_t$ is drawn for context.
  \textbf{Panel~D:} The exact decomposition of raw IU into its structural (artifact), genuine and
  idiosyncratic components: the structural share averages $74\%$ over the run and lies between
  $44\%$ and $96\%$ in nine periods out of ten.
  The series of Panels~B--C are standardized. Script: \texttt{fig14\_mc\_decomposition.py}.}
  \label{fig:mc_decomposition}
\end{figure}

In the data the latent objects are never observed, so the case for a corrected measure
cannot be closed there. A simulation can close it, because the truth is set by construction.
The economy generated here has a latent inflation target that drifts, genuine belief
imprecision $u_t$ and directional risk $\delta_t$ following independent autoregressive
processes, and a panel of thirty heterogeneous forecasters who observe the economy and
report histograms on a grid, so that every measure in this paper can be computed from the
simulated reports exactly as it is computed from the survey. Five hundred replications are
drawn. Table~\ref{tab:mc_simulation_summary} reports the results and Figure~\ref{fig:mc_decomposition}
draws one long run.

\paragraph{Do the measures recover what they are meant to?}
Normalized Uncertainty tracks genuine imprecision better than raw dispersion does --- $R^2$
of $0.83$ against $0.58$ across replications --- and carries far less of the distance
artifact: regressed on the distance itself, the corrected measure returns $0.09$ where the
raw measure returns $0.44$, a contamination four and a half times smaller. Asymmetry
Coherence recovers the true directional state better than raw asymmetry, $0.94$ against
$0.78$. Neither result is surprising given the construction; what the simulation adds is
that the gap is this wide at realistic noise levels rather than only in the limit.

\paragraph{What the contamination does to inference.}
The sharper diagnostic runs the failure the correction is meant to prevent. In a
data-generating process where an outcome depends on the distance of expectations from the
anchor but not at all on genuine uncertainty, raw dispersion enters a regression of that
outcome significantly in $58\,\%$ of replications with heteroskedasticity- and
autocorrelation-robust standard errors: the measure carries the distance, and the distance
carries the outcome. The corrected measure cuts that to $24\,\%$ --- better, and still far
from innocent. Adding the distance as a control collapses the rejection rate to its nominal
level for both, $4.6\,\%$ each, which identifies the channel exactly: the spurious finding
is the level, travelling under the name of uncertainty. Two things follow for practice. A
corrected measure is not a substitute for controlling for the level when the level is what
one is worried about; and the reason to prefer the corrected measure is not that it is
immune, but that what survives in it is a smaller and cleaner thing.

\paragraph{How much does the calibration matter?}
Section~\ref{sec:nu_envelope} publishes two series because one number separates them, and the
simulation misspecifies that number on purpose. Setting the envelope's intercept or its
slope wrong by a factor of two leaves the corrected measure tracking genuine uncertainty at
$0.61$ to $0.94$, against $0.58$ for the raw measure, with contamination between $0.02$ and
$0.28$ against the raw $0.44$: the ranking survives, which is what licenses reporting a
fitted and a calibrated series side by side rather than defending one. The exception is
instructive. Misspecifying the \emph{anchor} --- putting $\mu^{*}$ a full point away from
the value beliefs are actually organized around --- drops the corrected measure to $0.44$,
below the raw measure it is meant to improve on. The calibration of the envelope is
forgiving; the choice of the reference point is not. That is a caution for every setting in
which the anchor is estimated rather than announced, which is the situation of the growth
measure in Section~\ref{sec:growth}.

\paragraph{The aggregator.}
A separate experiment varies the rule that combines the first and third moments and finds
that the coherence weight of~\eqref{eq:ac} does not change which direction is recovered ---
a weight in $[0,1]$ cannot flip a sign --- but calibrates the strength of the signal,
amplifying it on dates when the two moments agree and damping it when they do not: in the
simulation both the raw third moment and the index recover the direction of the latent
signal on every replication, while the index explains the latent directional state far
better than raw asymmetry does, $0.73$ against $0.59$. That is the property the index is
built for, and the one a simple average of the two components does not have.
    \section{Conclusion}
\label{sec:conclusion}

This paper shows that higher moments of survey density forecasts should not be interpreted independently of their first moment. In inflation surveys, raw inflation dispersion rises predictably when expected inflation moves above the target, but is flat below it; raw asymmetry, measured from sparse histograms, is too noisy to serve as a stand-alone measure of directional risk. We therefore construct two corrected statistics. Normalized Uncertainty (NU) scales reported dispersion by the variance--distance envelope implied by the forecast location. Asymmetry Coherence (AC) combines asymmetry with the deviation of the central forecast from the target and discounts configurations in which the two signals conflict.

The empirical (ECB Survey of Professional Forecasters) and simulation exercises indicate that these corrections recover distinct objects. NU is less mechanically related to the inflation level and more closely tracks latent belief imprecision in the simulation, while AC better recovers the latent directional state than raw asymmetry. The extensions to growth forecasts and the US Survey of Professional Forecasters also clarify the conditions for portability: the reference point, treatment of tails, and survey-specific measurement conventions are substantive features of the measure. In particular, a change in the histogram grid can alter measured dispersion without any change in beliefs. Across the common euro-area and US sample, realized and expected inflation co-move closely, whereas NU does not, consistent with it capturing uncertainty distinct from the level of inflation.

These measures are measurement corrections reducing the risk of interpreting deviations from the anchor as uncertainty about inflation, or noisy histogram asymmetry as the balance of risks. The resulting measures provide inputs for the analysis of monetary-policy tolerance in \citet{vansteenberghe2026tolerable} and for the credit-supply application in \citet{vansteenberghe2025monetary}.        

    \newpage
    \bibliographystyle{aea}
    \bibliography{literature}


\begin{figure}[tbp]
  \centering
  \includegraphics[width=\linewidth]{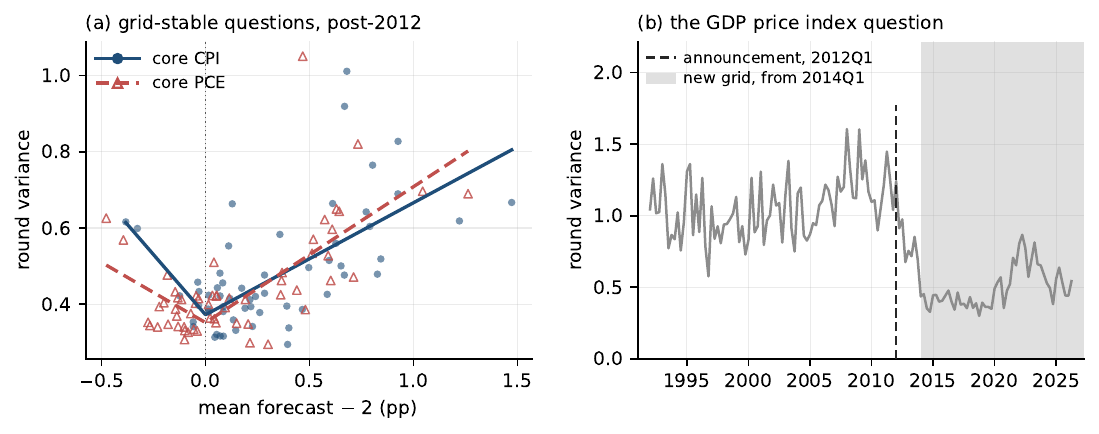}
  \caption{\textbf{The envelope in the US survey.} \emph{Panel (a):} the round-level total variance
  of the reported densities against the signed distance of the round's mean forecast from $2\,\%$, for
  the two grid-stable core questions over the fifty-eight rounds after the January 2012 announcement,
  with the two-arm fit of Table~\ref{tab:us_arms}; core CPI is drawn with filled circles and a solid
  fit, core PCE with open triangles and a dashed fit. \emph{Panel (b):} the same variance for the GDP
  price index question, round by round, with the January 2012 announcement dashed and the window of the
  new bin widths shaded.}
  \label{fig:us_arms}
  \small
  \textit{Note:} panel (b) begins in 1992Q1. The question was suspended after 1980Q2 and resumed then,
  and of the twenty-one rounds before the suspension only one has a mean forecast inside the window the
  estimates are formed on, so the earlier record is not drawn. Mean measured variance falls from $1.04$
  to $0.50$ across the change of grid.
  Script: \texttt{fig12\_us\_arms.py}.
\end{figure}

\begin{table}[tbp]\centering
\caption{The variance--distance envelope, euro-area panel}\label{tab:arms}
\begin{tabular}{lccccc}
\toprule
 & $a$ & $b_-$ & $b_+$ & $R^2$ & $n$ \\
\midrule
\multicolumn{6}{l}{\emph{average individual variance $W_t$: the envelope the measure divides by}}\\
two arms, all rounds & $0.390$ & $-0.007$ & $0.853$ & $0.697$ & $110$ \\
 & ($0.064$) & ($0.098$) & ($0.072$) & & \\
two arms, rounds before 2020 & $0.288$ & $0.127$ & $0.431$ & $0.177$ & $84$ \\
 & ($0.033$) & ($0.049$) & ($0.140$) & & \\
\addlinespace
\multicolumn{6}{l}{\emph{the same rounds: disagreement, and the total}}\\
disagreement $D_t$ & $0.036$ & $0.115$ & $1.028$ & $0.719$ & $110$ \\
 & ($0.027$) & ($0.047$) & ($0.115$) & & \\
total $T_t=W_t+D_t$ & $0.426$ & $0.108$ & $1.881$ & $0.788$ & $110$ \\
 & ($0.074$) & ($0.116$) & ($0.138$) & & \\
\bottomrule
\end{tabular}
\vspace{0.3em}
\parbox{0.95\linewidth}{\footnotesize \textit{Notes:} round-level regressions of each variance on
the signed distance of the consensus forecast (the cross-forecaster mean) from the $2\,\%$
target, in the two-arm form of~\eqref{eq:vstruct}. $W_t$ is the round mean
of the individual density variances, $D_t$ the cross-forecaster variance of their means.
ECB-SPF one-year-ahead inflation densities, 1999Q1 to the latest round; rounds whose consensus
lies outside $[-1,5]$ are excluded, which removes one round, 2022Q4 (consensus
$5.01$). Newey--West standard errors at four lags in parentheses. Wald tests on the two-arm
fit of $W_t$: $b_-=0$ gives $\chi^2_1=0.005$ ($p=0.94$); $b_-=b_+$ gives $\chi^2_1=91.9$
($p<0.001$). $W_t$ is 73 per cent of $T_t$. The measure uses the ratio of the $W_t$ fit
through 2026Q2, $b_+/a=2.23$ ($n=109$, delta-method standard error 0.49); on these $110$ rounds it
is $2.19$. Fitted on the rounds before 2020 the same ratio is $1.50$ (0.63): over those rounds the
consensus gap ran over $[-1.27,+0.49]$ and only 9 rounds sat above target, against
$[-1.01,+1.76]$ from 2020Q1, so the upper arm is barely identified there. The two periods do not share
one envelope ($\chi^2_3=380$, $p<0.001$, the shift being the intercept rather than the upper arm),
while the difference between the two ratios alone is within one standard error of zero. A fitted ratio is
therefore a dated object, tied to the range of distances the sample happens to contain; the calibration
$r_+=1$ of~\eqref{eq:nu_cal} is not. Script: \texttt{tab01\_arms.py}.}
\end{table}

\begin{table}[tbp]\centering
\caption{Moments of the reported densities: correlations}\label{tab:momentscorr}
\begin{tabular}{lccc}
\toprule
\multicolumn{4}{l}{\emph{across survey rounds} ($n=111$)}\\
 & Mean & CV & Bowley \\
\midrule
Mean & $1.00$ & $-0.49$ & $0.36$ \\
CV & $-0.49$ & $1.00$ & $-0.37$ \\
Skew & $0.36$ & $-0.37$ & $1.00$ \\
\addlinespace
\multicolumn{4}{l}{\emph{across individual forecaster-rounds} ($n=4624$)}\\
 & Mean & CV & Bowley \\
\midrule
Mean & $1.00$ & $-0.23$ & $0.11$ \\
CV & $-0.23$ & $1.00$ & $-0.03$ \\
Skew & $0.11$ & $-0.03$ & $1.00$ \\
\bottomrule
\end{tabular}
\vspace{0.3em}
\parbox{0.95\linewidth}{\footnotesize \textit{Notes:} Bowley's skewness as defined in
Section~\ref{sec:moments}; CV is a density's standard deviation over its mean. The upper panel
averages each moment within a survey round before correlating; the lower panel correlates the
forecaster-round observations themselves. The mean--asymmetry relation is positive at both
levels and an order of magnitude weaker at the level where the third moment is measured. Script: \texttt{tab02\_moments\_correlations.py}.}
\end{table}

\begin{table}[tbp]\centering
\caption{Asymmetry against the signed distance from target}\label{tab:skew_dev_regs}
\begin{tabular}{lcccc}
\toprule
 & (1) & (2) & (3) & (4) \\
\midrule
$\mu_{i,t}-\mu^{*}$ & $0.0128$ & $0.0147$ & $0.0126$ & $0.0103$ \\
                    & $(0.0018)$ & $(0.0035)$ & $(0.0028)$ & $(0.0045)$ \\
$t$                 & $7.01$ & $4.24$ & $4.43$ & $2.31$ \\
\midrule
Observations        & $4{,}624$ & $4{,}624$ & $4{,}549$ & $4{,}549$ \\
$R^2$               & $0.011$ & $0.056$ & $0.006$ & $0.053$ \\
Survey-round effects & No & Yes & No & Yes \\
Trimmed $|d_{i,t}|\leq 2$ & No & No & Yes & Yes \\
\bottomrule
\end{tabular}
\vspace{0.3em}
\parbox{0.95\linewidth}{\footnotesize \textit{Notes:} ordinary least squares of individual
Bowley skewness on the signed deviation of that forecaster's density mean from the $2\,\%$
target, ECB-SPF one-year-ahead inflation densities, 1999Q1 to the latest round. Standard
errors clustered by survey round in parentheses. Columns (2) and (4) add a full set of
survey-round dummies, so the slope is identified from disagreement within a round. A positive
and significant coefficient says that asymmetry is coherent on average, $c>0$ in the notation
of Section~\ref{sec:distance_asymmetry}. The small $R^2$ is the noise the index of
Section~\ref{sec:AC_measure} is built to handle. Script: \texttt{tab03\_skewness\_deviation.py}.}
\end{table}

\begin{table}[tbp]\centering
\caption{Inflation and growth uncertainty}\label{tab:ngu_orth}
\begin{tabular}{lc}
\toprule
\multicolumn{2}{l}{\emph{the matched panel}}\\
forecaster-rounds with both densities, through 2026Q2 & $4{,}327$ \\
\hspace{1em} through the latest round & $4{,}350$ \\
forecasters carrying their own slope ($\geq 10$ matched rounds) & $86$ \\
forecasters taking the within-forecaster slope & $21$ \\
\addlinespace
\multicolumn{2}{l}{\emph{correlation between the two round series}}\\
full sample, 1999Q1--2026Q3 & $+0.66$ \\
\addlinespace
\multicolumn{2}{l}{\emph{agreement with the text-based uncertainty index}}\\
raw growth standard deviation & $0.60$ \\
$\mathrm{NGU}$ & $0.72$ \\
\addlinespace
\multicolumn{2}{l}{\emph{the orthogonalized components}}\\
correlation of $\mathrm{NGU}\perp\mathrm{NU}$ built on the two NU series & $0.99$ \\
\bottomrule
\end{tabular}
\vspace{0.3em}
\parbox{0.95\linewidth}{\footnotesize \textit{Notes:} the matched panel is the set of
forecaster-rounds reporting both an inflation and a GDP-growth density. The purge is
\eqref{eq:orth}, run forecaster by forecaster. Script: \texttt{tab04\_ngu\_orthogonalization.py}.}
\end{table}

\begin{table}[tbp]\centering
\caption{The envelope in the US survey, by density question}\label{tab:us_arms}
\begin{tabular}{lccccc}
\toprule
question / window & $n$ & $R^2$ sym. & $R^2$ arms & $b_-$ $(t)$ & $b_+$ $(t)$ \\
\midrule
core PCE, 2007--2026        & $78$ & $0.489$ & $0.493$ & $+0.414$ $(7.3)$ & $+0.364$ $(7.7)$ \\
\hspace{1em} before 2012    & $20$ & $0.583$ & $0.672$ & $+0.348$ $(3.0)$ & $-0.168$ $(-0.7)$ \\
\hspace{1em} after 2012     & $58$ & $0.472$ & $0.473$ & $+0.316$ $(2.3)$ & $+0.357$ $(8.0)$ \\
\addlinespace
core CPI, 2007--2026        & $78$ & $0.310$ & $0.398$ & $+0.574$ $(8.9)$ & $+0.268$ $(5.1)$ \\
\hspace{1em} before 2012    & $20$ & $0.042$ & $0.708$ & $+0.183$ $(1.3)$ & $-0.630$ $(-2.8)$ \\
\hspace{1em} after 2012     & $58$ & $0.410$ & $0.433$ & $+0.631$ $(7.6)$ & $+0.294$ $(6.7)$ \\
\addlinespace
GDP price index, old grid to 2011 & $81$ & $0.046$ & $0.051$ & $+0.025$ $(0.1)$ & $-0.109$ $(-0.9)$ \\
\hspace{1em} new grid, 2014--  & $50$ & $0.553$ & $0.558$ & $+0.194$ $(1.6)$ & $+0.273$ $(6.0)$ \\
\addlinespace
euro area, for comparison   & $109$ & $0.307$ & $0.789$ & $+0.117$ $(1.0)$ & $+1.882$ $(13.7)$ \\
\midrule
\multicolumn{6}{l}{\emph{the two events run against each other} (GDP price index, individual level, $|d|\leq 2$)}\\
\multicolumn{4}{l}{change of grid, 2014 $\times\,|d|$} & \multicolumn{2}{c}{$-0.413$ $(-1.4)$} \\
\multicolumn{4}{l}{announcement, 2012 $\times\,|d|$} & \multicolumn{2}{c}{$+0.185$ $(+0.6)$} \\
\midrule
\multicolumn{6}{l}{\emph{the within-source ratio} $b_+/a$, individual level}\\
\multicolumn{4}{l}{euro area} & \multicolumn{2}{c}{$2.61$} \\
\multicolumn{4}{l}{US core CPI, after 2012} & \multicolumn{2}{c}{$0.20$} \\
\multicolumn{4}{l}{US core PCE, after 2012} & \multicolumn{2}{c}{$0.12$} \\
\bottomrule
\end{tabular}
\vspace{0.3em}
\parbox{0.95\linewidth}{\footnotesize \textit{Notes:} round-level regressions of the total
density variance on the two arms of the signed distance from $2\,\%$, Newey--West standard
errors at four lags, except where stated. The core questions have used one grid since 2007;
the GDP price index question halved its bin widths in 2014Q1 and is therefore split by grid,
never across it; its old-grid row stops at the announcement, and the eight old-grid rounds of
2012--13 are in the script's results file. The race at the foot of the table is the specification in which the published
pre/post-2012 break was estimated: the interaction that carries the break is the change of
grid, not the announcement. Slopes are never compared across sources, only shapes and the
within-source ratio. Script: \texttt{tab05\_us\_arms.py}.}
\end{table}

\begin{table}[tbp]
\centering
\small
\caption{Monte Carlo simulation: raw versus corrected moments}\label{tab:mc_simulation_summary}
\begin{tabular}{@{}llcc@{}}
\toprule
\textbf{Strategy} & \textbf{Metric} & \textbf{Raw} & \textbf{Corrected} \\
\midrule
\multicolumn{4}{@{}l}{\emph{Panel A: Normalized Uncertainty}} \\[2pt]
A.\ Oracle $R^2$: tracking $u_t$ & $R^2(\cdot,\, u_t)$ & 0.581 & \textbf{0.829} \\
A.\ Artifact exposure: tracking $d_t$ & $R^2(\cdot,\, d_t)$ & 0.437 & \textbf{0.094} \\
B.\ Spurious rejection (HAC, outcome $\sim$ measure) & \% reject at 5\% & 57.6\% & 23.6\% \\
B.\ Controlled (HAC, outcome $\sim$ measure $+$ $d$) & \% reject at 5\% & 4.6\% & \textbf{4.6\%} \\
\addlinespace[4pt]
\multicolumn{4}{@{}l}{\emph{Panel B: Asymmetry Coherence}} \\[2pt]
A.\ Coherent signal recovery & $R^2(\cdot,\, \mathrm{AC}^{nf})$ & 0.780 & \textbf{0.941} \\
\bottomrule
\end{tabular}
\vspace{0.2em}
\begin{minipage}{0.97\linewidth}\footnotesize
\textit{Notes:} Monte Carlo simulation with 500 replications, $T=$200 periods, $N=$30 forecasters. The DGP is an extended UCSV model with a latent inflation target (mean-reverting with state-dependent drift, $\kappa_\theta=$0.02), AR(1) genuine uncertainty $u_t$, and AR(1) directional risk $\delta_t$. Raw: cross-sectional mean of individual variances (IU) or raw asymmetry (Bowley's formula). Corrected: NU $= \sqrt{\mathrm{Var}}/\sqrt{a + b\,|{\mu-\mu^*}|}$ or AC $= [(\tilde{Q}+\tilde{A})/2]\cdot[(1+\tilde{Q}\tilde{A})/2]$. $\mathrm{AC}^{nf}$: noise-free AC computed from structural and genuine Bowley components (without incoherent noise). Strategy~B uses Newey--West (HAC) standard errors; Strategy~D uses expanding-window (real-time) AC normalisation with no look-ahead. Bold: the measure with better performance. Full results in Section~\ref{sec:simulation}. Script: \texttt{tab06\_mc\_summary.py}.
\end{minipage}
\end{table}

\end{document}